\documentclass[epj]{svjour}
\usepackage{amssymb}
\usepackage{graphicx}
\usepackage[latin1]{inputenc}
\usepackage{amsmath}

\usepackage[sort&compress]{natbib}
\bibpunct[, ]{[}{]}{,}{n}{,}{,}
\usepackage{mciteplus}

\newcommand{\nuc}[2]{\ensuremath{^{#1}}#2}
\newcommand{\sys}[4]{\ensuremath{^{#1}}#2+\ensuremath{^{#3}}#4}
\newcommand{\AM}{$A\,$MeV}

\begin{document}

\title {Isospin effects and symmetry energy studies with INDRA}

\author{
G.~Ademard\inst{1}
\and B.~Borderie\inst{1}
\and A.~Chbihi\inst{2}
\and O.~Lopez\inst{3}
\and P.~Napolitani\inst{1}
\and M. F.~Rivet\inst{1}\mail{rivet@ipno.in2p3.fr}
\and M.~Boisjoli\inst{2,4}
\and E.~Bonnet\inst{2}
\and R.~Bougault\inst{3}
%\and F.~Gagnon-Moisan\inst{1,4}\thanks{Present address: PSI, 5234 Villingen,
%Switzerland}
\and J. D.~Frankland\inst{2}
\and E.~Galichet\inst{1,5}
\and D.~Guinet\inst{6}
\and M.~Kabtoul\inst{3}
\and G.~Lehaut\inst{3}
\and P.~Lautesse\inst{7}
\and M.~La~Commara\inst{8}
\and N.~Le~Neindre\inst{3}
\and P.~Marini\inst{2}
\and M.~P\^arlog\inst{3}
\and P.~Paw{\l}owski\inst{8}
\and E.~Rosato\inst{9}
\and R.~Roy\inst{4}
\and E.~Spadaccini\inst{9}
\and E.~Vient\inst{3}
\and M.~Vigilante\inst{9}
\and J. P. Wieleczko\inst{2}\\
\and (INDRA Collaboration)
}

\institute{
Institut de Physique Nucl\'eaire, CNRS-IN2P3 and Universit\'e Paris-Sud 11,
F-91406 Orsay cedex, France
\and GANIL, CEA-DSM/CNRS-IN2P3, B.P.~5027, F-14076 Caen cedex, France.
\and LPC Caen, ENSICAEN, Universit\'e de Caen, CNRS-IN2P3, F-14050 Caen cedex, France.
\and Universit\'e Laval, Qu\'ebec, G1V 0A6 Canada.
\and Conservatoire National des Arts et M\'etiers, F-75141 Paris cedex 03,
France.
\and Institut de Physique Nucl\'eaire, UCBL, Universit\'e de Lyon, CNRS-IN2P3,
F-69622 Villeurbanne cedex, France.
\and S2HEP (EA4148), UCBL/ENSL, Universit\'e de Lyon, Villeurbanne, France.
\and IFJ-PAN, 31-342 Krak\'ow, Poland
\and Dipartimento di Scienze Fisiche e Sezione INFN, Universit\'a
di Napoli ``Federico II'', I80126 Napoli, Italy.
}

\date{\today}

\abstract{The equation of state of asymmetric nuclear matter is still
controversial, as predictions at subsaturation as well as above normal
density widely diverge. We discuss several experimental results measured in
heavy-ion collisions with the INDRA array in the incident energy range
5-80~\AM{}. In particular an estimate of the density dependence of the 
symmetry energy is derived from isospin diffusion results compared
with a transport code: the potential part of the symmetry energy linearly
increases with the density. We demonstrate that isospin equilibrium is
reached in mid-central collisions for the two reactions Ni+Au at 52~\AM{}
and Xe+Sn at 32~\AM{}.
New possible variables and an improved modelization to investigate symmetry 
energy are  discussed. 
}

 
\maketitle

%%%%%%%%%%%%%%%%%%%%%%%%%%%%%%%%%%%%%%%%%%%%%%%%%%%%%%%%%%%%%%%%%%%%%
\section{Introduction\label{intro}}
%%%%%%%%%%%%%%%%%%%%%%%%%%%%%%%%%%%%%%%%%%%%%%%%%%%%%%%%%%%%%%%%%%%%%
Heavy-ion collisions are used as an experimental probe for isospin effects
related to the equation of state of isospin-asymmetric nuclear matter.
At the same time, microscopic models are worked out in order to
incorporate and test isospin effects in the description of nuclear processes.
The physical picture and the corresponding modelling requirements depend on the 
range of incident energies they apply to.
Several different forms are suggested for the isospin contribution to the
equation of state, in particular as a consequence of many alternative 
descriptions  of the density dependence of the symmetry energy~\cite{Fuc06}.
The difference between these forms stands out dramatically when the saturation
density is exceeded; since such condition can be achieved at large bombarding 
energies, the corresponding modelling approach should be built by taking 
into account, in the isovector channel, effects like the momentum dependence 
and the hadron effective mass splitting~\cite{Dit09}.

The route taken by the INDRA collaboration is to investigate
the subsaturation-density domain, explored at Fermi energies.
The specific interest of this choice, although the sensitivity to the density 
dependence of the symmetry energy is reduced (implying to take into account 
also finer effects like the secondary decay), is that in this energy regime 
isospin effects can be searched in relation with transport observables, 
like diffusion and migration processes~\cite{Bar05,Riz08,Nap10},
or with cluster-correlation properties and fragment observables,
like phase-transition features~\cite{Cho00,*Cho03},
bimodalities~\cite{I61-Pic06,I72-Bon09}, 
spinodal effects and distillation processes~\cite{Duc07}.
This is the physical landscape addressed by INDRA.

The 4$\pi$ multidetector INDRA, which is described in detail
in~\cite{I3-Pou95,I5-Pou96}, was thus used to reveal N/Z effects in heavy-ion
reactions connected to the knowledge of symmetry energy entering the
asymmetric nuclear equation of state~\cite{Bar05,WCI06,Bao08}. In terms of
detection INDRA possesses excellent performance: geometrical efficiency of
90\%, rather low detection and identification thresholds
(see Fig. 1 of ref.~\cite{I28-Fra01}), accurate charged particle
and fragment identifications, energy measurements with an accuracy of 4\%.
Further details can be found in Refs.~\cite{I14-Tab99,I33-Par02,I34-Par02}.
The global quality of detection was then used to, in particular, perform
data selections and reconstruct excited nuclear systems to compare with 
simulations over large ranges of impact parameters. 
\par
The paper is organized as follows.  Results obtained for central collisions
are evidenced in section~\ref{sect3}  and a new variable to investigate 
symmetry energy effects is proposed. In section~\ref{sect4}, 
semiperipheral collisions are discussed and evolutions of observables 
with the impact parameter are compared to theoretical simulations 
to derive information on the potential part of the symmetry energy term
from isospin diffusion. Average stopping for central collisions
with different N/Z entrance channels are presented in section~\ref{sect5}. 
Section~\ref{sect6} is devoted to studies in progress related to projectile 
fragmentation at Fermi energy, fusion reactions at low energies with 
different N/Z and comparison of data with a new transport code.
Some foreseen studies are described in section~\ref{sect7}. Conclusions are 
drawn in section~\ref{sect8}. Finally in the appendix we gather the 
different experimental selections made for the presented data, some of them
being specific to the INDRA array.

%%%%%%%%%%%%%%%%%%%%%%%%%%%%%%%%%%%%%%%%%%%%%%%%%%%%%%%%%%%%%%%%%%%%%%%%%%%%
\section{Isospin effects in central collisions\label{sect3}}
%%%%%%%%%%%%%%%%%%%%%%%%%%%%%%%%%%%%%%%%%%%%%%%%%%%%%%%%%%%%%%%%%%%%%%%%%%%%

In this section we describe isospin effects observed in 
quasi-fusion reactions. First constraints on the symmetry term of the EOS
could be obtained by studying nuclear reactions with judiciously chosen 
projectile-target 
couples. We bombarded \nuc{112,124}{Sn} targets with \nuc{124,136}{Xe} 
projectiles accelerated at 32 and 45~\AM{} by the GANIL facility; the four 
possible configurations were studied at 32\AM{} while 
\sys{124}{Xe}{124}{Sn} was not at 45\AM{}. The \sys{124}{Xe}{124}{Sn} 
and \sys{136}{Xe}{112}{Sn} systems have the same combined isospin (N/Z=1.385), 
while the isospin gradient between projectile and target is 1.5 times larger 
for the second one.
 
Central collisions  were selected as explained
in section~\ref{sect21}. Results concerning the LCP and fragment (Z$\geq$5)
multiplicities, as well as the centre of mass average fragment kinetic
energies, were published in~\cite{I77-Gag12}. However, as mentioned in that
reference, we collected a very large number of events; it thus becomes 
possible to extrapolate the values of different variables to those 
corresponding to a perfect detection of charges, Z$_{tot}$=104 (the
procedure is described in the appendix~\ref{sect21}).
This will facilitate the comparison with simulations, avoiding the 
filtering  of the  calculated events.  We observed that 
the widths of the multiplicity distributions are not modified
by the degree of completeness, whereas those of the Z$_{bound}$
variables decrease for more complete events. 
We verified that the charge and angular distributions of charged
products were not modified when strengthening the event completeness. 

\begin{figure}[htb]
\includegraphics[width=0.5\columnwidth]{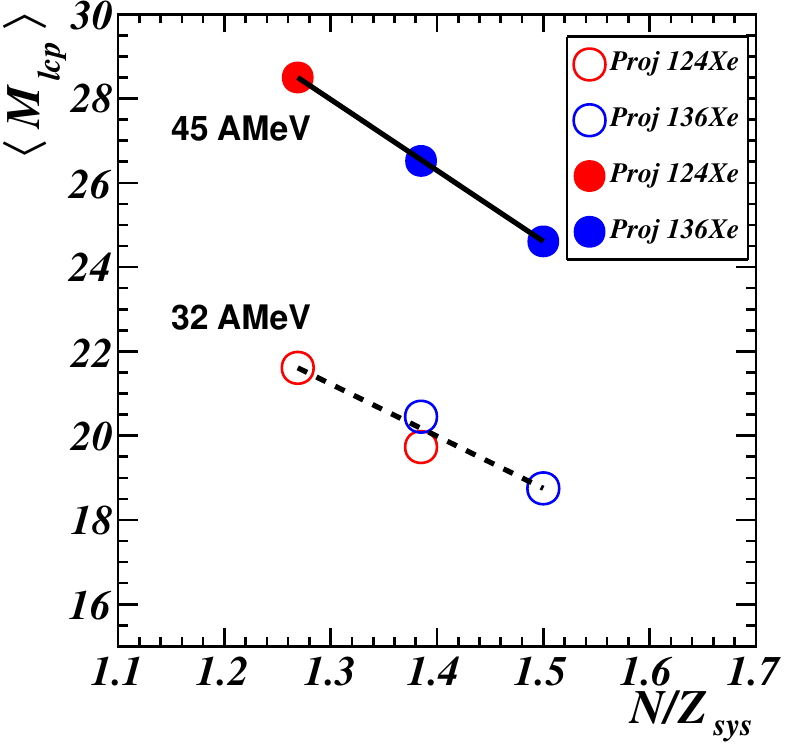}%
\includegraphics[width=0.5\columnwidth]{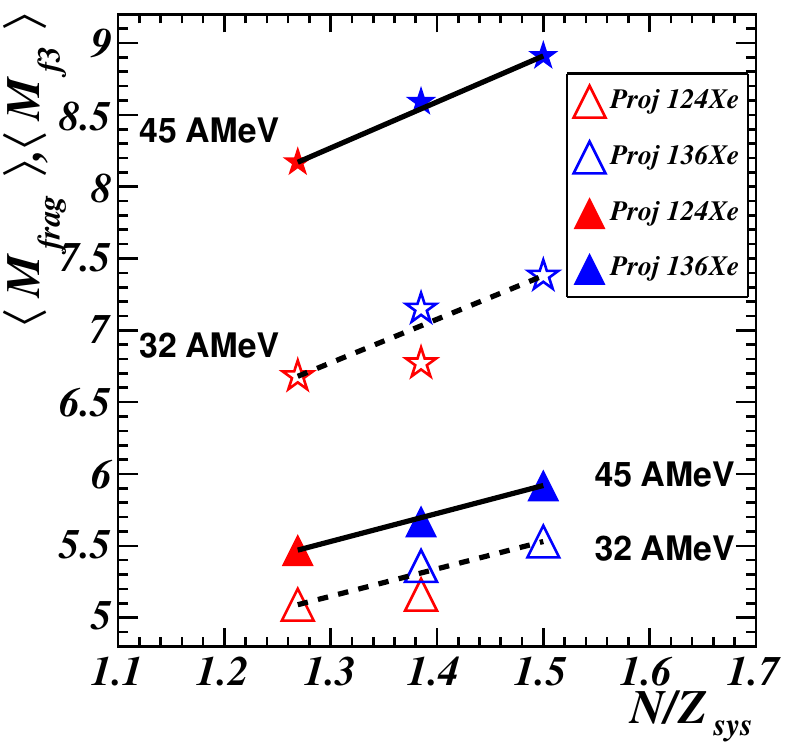}
\caption{(Colour on line) Left: evolution of the mean multiplicities of 
LCP measured for 
quasi-fusion reactions \emph{vs} the N/Z of the different Xe+Sn total 
systems at 32 and 45~\AM{}. Right: same for mean fragment multiplicities, 
$M_{f3}$ (Z$\geq$3, stars) 
and $M_{frag}$ (Z$\geq$5, triangles).}
\label{fig1}
\end{figure}

Obviously the multiplicities obtained with this procedure are larger 
than those published in~\cite{I77-Gag12}, but the trends observed as a 
function of the total system isospins are not modified, as displayed in 
fig.~\ref{fig1}.
%%%%%%%%%%%%%%%%%%%%%%%%%%%%%%%%%%%%%%%%%%%%%%%%%%%%%%%%%%%%%%%%%%%%%%%%%%%%%%
%%%%%%%%%%%%%%%%%%%%%%%%%%%%%%%%%%%%%%%%%%%%%%%%%%%%%%%%%%%%%%%%%%%%%%%%%%%%%%%%
$M_{lcp}$ linearly decreases with increasing isospin, with a steeper slope 
at the higher energy, whereas $M_{frag}$ linearly increases, with equal slopes 
at 32 and 45~\AM{}. At 32~\AM{}, we  measured the two mixed systems, 
\sys{124}{Xe}{124}{Sn} and \sys{136}{Xe}{112}{Sn}; 
we observe that both LCP and fragment multiplicities little depend on the 
entrance channel mass asymmetry.

In short we observed that in multifragmentation of quasi-fused systems, more 
fragments, and less LCP, are emitted when the system grows neutron-richer. 
In previous works the increase of fragment multiplicity was attributed to 
phase-space effects~\cite{Kun96}. We underlined in~\cite{I77-Gag12} that the 
10\% difference measured both at 32 and 45~\AM{} between the lightest and 
the heaviest systems correspond to the mass increase between them. It recalls 
the scaling law observed in~\cite{I12-Riv98} and attributed to volume 
instabilities as the origin of the multifragmentation phenomenon (spinodal 
decomposition). We tested this hypothesis by comparing our data to a 
stochastic transport code.

%%%%%%%%%%%%%%%%%%%%%%%%%%%%%%%%%%%%%%%%%%%%%%%%%%%%%%%%%%%%%%%%%%%%%
\subsection{The Stochastic Mean Field model \label{sect31}}
%%%%%%%%%%%%%%%%%%%%%%%%%%%%%%%%%%%%%%%%%%%%%%%%%%%%%%%%%%%%%%%%%%%%%
At Fermi energy and below an appropriate description of the density profile 
in phase space demands an accurate description of mean-field properties and 
a precise treatment of the Pauli blocking.
At these energies, even if the early topology of the system may 
 suggest a final fragmented configuration, transport mechanisms driven
by density gradients (isospin migration) or by isospin gradients 
(isospin diffusion) may still progress up to few hundred fm/$c$ and 
affect the freeze-out configuration. In order to apply efficiently to 
such  situation a one-body description should therefore be preferred,
as it is fully adapted to follow the bulk behaviours for a long 
interval of time, corresponding with the complete duration of the process.
In this framework, central and semicentral collisions are particularly 
interesting: they allow to explore the largest span in nuclear density and to
access very small values compatibles with the disassembling of the
system. According to experimental and theoretical studies~\cite{Cho04,Bor08},
at bombarding energies around 30AMeV, for a system of average size like Xe+Sn, 
the dynamics of the system should be largely determined
by spinodal behaviours: this is actually a consequence of the form of
the equation of state and of the properties of the mean field which
stands out when approaching the low-energy threshold of multifragmentation.

The above phenomenology is well described using the Boltzmann-Langevin (BL)
equation that describes the evolution of the semiclassical one-body 
distribution function $f({\bf r},{\bf p},t)$ submitted to the effective 
Hamiltonian $H[f]$ and the residual interaction; this latter, expressed 
in terms of the one-body distribution function $f$ includes the average 
Boltzmann collision integral $\bar{I}[f]$, and the fluctuating term 
$\delta I[f]$:
\begin{equation}
	\dot{f}	= \partial_t\,f - \left\{H[f],f\right\}
			= {\bar{I}[f]}+{\delta I[f]} \;.
\label{eq1}
\end{equation}
Various approximate treatments of the BL equation have been developed.
The Stochastic Mean Field model, SMF, uses the following 
method~\cite{Colon98}:
within the assumption of local thermal equilibrium, the stochastic term 
of Eq.~\ref{eq1}, $\delta I[f]$, builds kinetic equilibrium fluctuations
typical of a
Fermi gas, which are projected on density fluctuations in the coordinate 
space. The fluctuations so introduced are amplified by the
unstable mean field, and the dynamics is essentially driven by the
propagation of mean field instabilities~\cite{Col04}. 

We compared the results of this section and of the next one with the SMF
model. The free nucleon-nucleon cross section with its angular, energy 
and isospin dependence is used. We take a soft isoscalar equation of state, 
with a compressibility modulus K = 200 MeV. Two different
parameterizations of the symmetry energy are used; in the ``asystiff'' one 
the potential term increases linearly with the density whereas the ``asysoft''
one is taken from the SKM* Skyrme formulation~\cite{Bar02}.

The symmetry term of the EOS, $E_{sym}/A$, is often parameterized as
\begin{equation} \label{Esym_gam}
\frac{E_{sym}}{A}(\rho)=\frac{C_{s,k}}{2} 
 (\frac{\rho}{\rho_0})^{2/3}+ 
\frac{C_{s,p}}{2} (\frac{\rho}{\rho _0})^{\gamma} 
\end{equation}
where $\rho_0$ is the nuclear saturation density. The first term is kinetic,
coming from Pauli correlations; the second term is the potential part, that
carries the isovector properties of the effective nuclear interaction.
The value of the $\gamma$ exponent, valid close to $\rho_0$,
determines whether the equation is 
``asystiff''($\gamma \geq$1, potential term continuously increasing with
$\rho$) or ``asysoft'' ($\gamma <$1, potential term presenting a maximum 
between $\rho_0$ and 2$\rho_0$). 

A second order expansion around $\rho_0$ of the potential term of the 
symmetry energy reads:  
\begin{equation} \label{Esym_L}
\frac{E_{sym}}{A}(\rho)=S_0 + \frac{L}{3} (\frac{\rho-\rho_0}{\rho_0})
 +\frac{K_{sym}}{18}(\frac{\rho-\rho_0}{\rho_0})^2
 \end{equation}
In the SMF framework values of L$\geq$75 correspond to an asystiff
$E_{sym}$~\cite{Bar05}.

At any step of the calculation fragments are recognized by applying a 
coalescence procedure to the 
one-body density, connecting nearby cells in which the density is larger 
than a cut-off value, taken equal to $\rho_0/$5 (``liquid phase''). 
We have shown in~\cite{I29-Fra01} that, at 300~fm/$c$, the fragment
multiplicity is independent of the exact value of the cut-off density.  
The remaining early emitted nucleons constitute the ``gas phase''. 
The fragment phase space configuration at 300 fm/$c$ is injected in the 
SIMON code~\cite{Dur92} which performs the secondary decay during the
propagation of all products under the Coulomb field, thus preserving 
space-time correlations. 

%%%%%%%%%%%%%%%%%%%%%%%%%%%%%%%%%%%%%%%%%%%%%%%%%%%%%%%%%%%%%%%%%%%%%
\subsection{Multiplicities and $Z_{bound}$ from SMF \label{sect32}}
%%%%%%%%%%%%%%%%%%%%%%%%%%%%%%%%%%%%%%%%%%%%%%%%%%%%%%%%%%%%%%%%%%%%%%%%%%%%%%
%%%%%%%%%%%%%%%%%%%%%%%%%%%%%%%%%%%%%%%%%%%%%%%%%%%%%%%%%%%%%%%%%%%%%
\begin{figure}[htb]
\includegraphics[width=0.9\columnwidth]{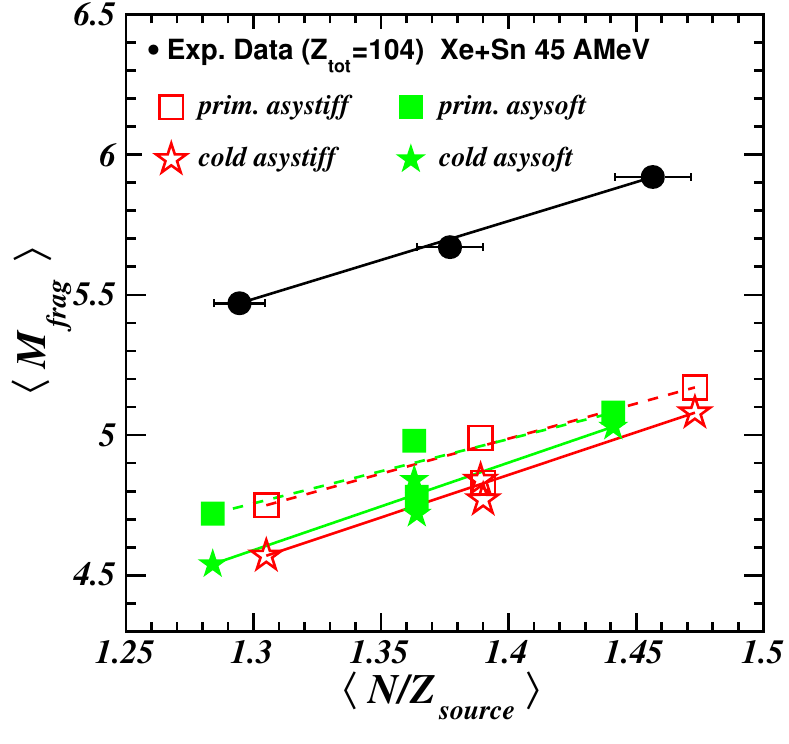}
\caption{ (Colour on line) Average fragment (Z$\geq$5) multiplicities, 
for the different 
systems  at 45~\AM{}: experimental values extrapolated to $Z_{tot}$=104 
(circles, the line is to guide the eye); calculated values for primary 
(squares) and cold fragments (stars), are plotted vs the source N/Z. 
Asystiff EOS: open symbols, asysoft EOS: filled symbols. In all cases
statistical error bars are smaller than the symbol sizes.}
\label{fig2}
\end{figure}
%%%%%%%%%%%%%%%%%%%%%%%%%%%%%%%%%%%%%%%%%%%%%%%%%%%%%%%%%%%%%%%%%%%%%%%%%%%%%%%%
For head-on collisions between Xe and Sn  at 45~\AM{} SMF predicts the
formation of a single source which subsequently breaks into several 
fragments. The fragment excitation energy at t=300~fm/$c$ is around 
3.3~\AM{}, in good agreement with experimental
evaluations~\cite{I39-Hud03,I66-Pia08}; we also verified that the
experimental charge distributions are reasonably accounted for by the
simulation. Conversely at 32~\AM{} the systems 
do not multifragment and lead to an evaporation residue,
because the treatment of the fluctuation mechanism is only described in average.
An improvement have therefore to be worked out on this aspect (see
section~\ref{sect63}).

%%%%%%%%%%%%%%%%%%%%%%%%%%%%%%%%%%%%%%%%%%%%%%%%%%%%%%%%%%%%%%%%%%%%%%%%%%%%%%
\begin{figure}[htb]
\includegraphics[width=0.5\columnwidth]{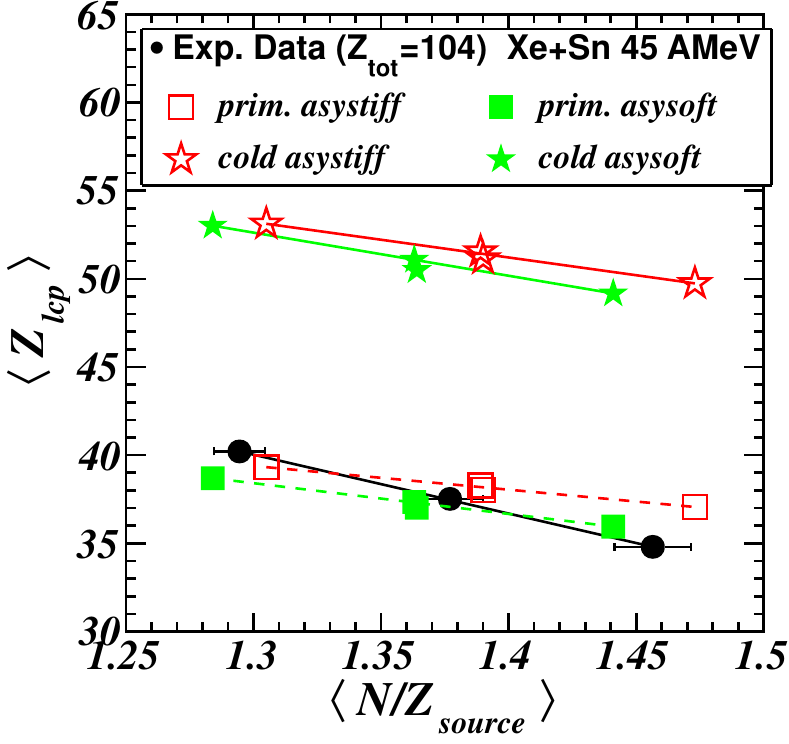}%
\includegraphics[width=0.5\columnwidth]{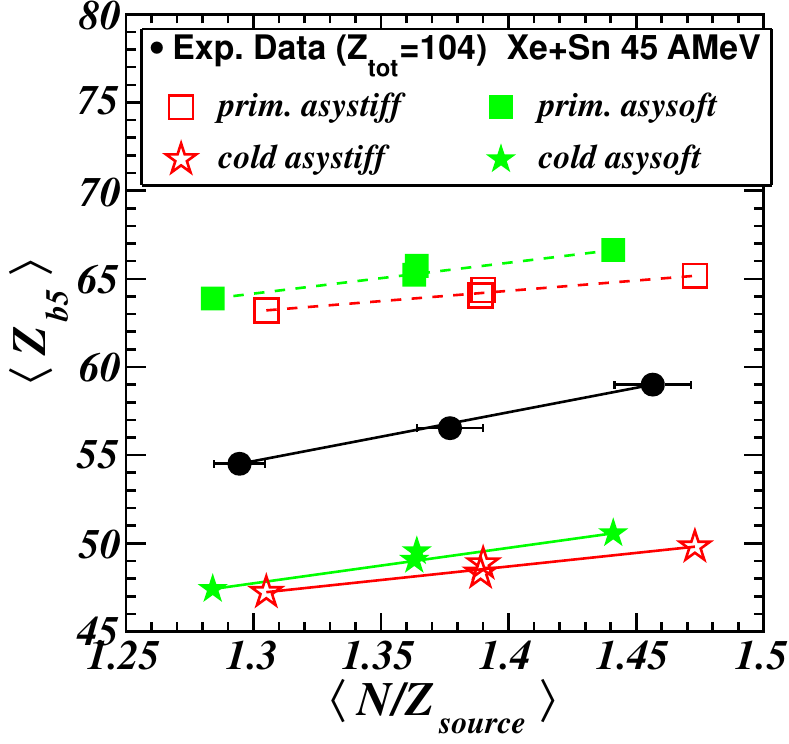}
\caption{(Colour on line) Left: evolution of the mean charge bound in 
LCP measured for 
quasi-fusion reactions \emph{vs} the N/Z of the different Xe+Sn total 
systems at 45~\AM{}. Right: same for mean charge bound in fragments, $Z_{b5}$,
Symbols as in fig.~\ref{fig2}.}
\label{fig3}
\end{figure}
%%%%%%%%%%%%%%%%%%%%%%%%%%%%%%%%%%%%%%%%%%%%%%%%%%%%%%%%%%%%%%%%%%%%%%%%%%%%%%%%
We present in figure~\ref{fig2} the measured and calculated
fragment multiplicities at 45~\AM{}. Results from the simulations are plotted 
at the N/Z value of the fragmenting source, at t=120~fm/c, 
after preequilibrium emission. Note that the N/Z range is reduced with 
respect to that covered by the initial systems. Indeed for both EOS the 
n/p ratio of preequilibrium is larger than that of the total system for
\sys{136}{Xe}{124}{Sn}, whereas it is smaller for \sys{124}{Xe}{112}{Sn};
consequently the source N/Z varies in reverse direction. For the mixed
systems, n/p and the source N/Z are identical to the system one in the
asystiff case. In all cases preequilibrium emission is more neutron-rich -
thus the source less neutron-rich - in the asysoft case. In figures 
\ref{fig2} and \ref{fig3}
experimental data are drawn at the average N/Z of the asysoft and
asystiff sources. We show only $M_{frag}$ and not $M_{f3}$ because the
former is little modified by the cooling stage, as appears from the
figure, comparing squares and stars, whereas the production of Li and Be is
increases during cooling. The difference between primary and final
multiplicities is even smaller for the neutron-rich system.  We note that
the simulations  lead to primary fragment multiplicities increasing with 
the neutron richness of the source: the observed effect comes from symmetry
energy, and is only slightly accentuated by evaporation. 
However the calculated final multiplicitie are smaller than the experimental 
one by about 1 unit (15\%). To get a deeper insight into the phenomenon, we
turn to $Z_{bound}$ variables. Indeed while the LCP multiplicity does not
make sense in the simulation, because essentially nucleons are produced in
the preequilibrium and multifragmentation phases, the charge bound in LCP is
meaningful. In figure~\ref{fig3} we present the evolution of $Z_{lcp}$ and 
$Z_{b5}$ with the source N/Z. We observe firstly that the experimental trend
is reproduced by the simulation already before de-excitation, confirming
that we are dealing with a symmetry energy effect. $Z_{lcp}$ at that stage
is already equal to the measured one, which indicates that the preequilibrium
emission is too abundant in the model. And indeed when we compare the data
with the final simulated values, we have too much charge bound in lcp and
not enough in fragments. 

The  difference between  the measured $Z_{b5}$ values for the 
neutron-richer and the 
neutron-poorer systems is about 5.5 units (9.8\%). It is smaller in the
simulation, 2.6 units (5.3\%) in the asystiff case and 3.2 (6.5\%) with 
the asysoft EOS. This observation, although it does not allow to make a
choice between the asy-EOS, indicates that the $Z_{b5}$ variable is more 
promising than the fragment multiplicity for constraining the EOS.

Another observable shown in~\cite{I77-Gag12} is the centre of mass average
kinetic energy of the fragments. At both incident energies we observed that
for each Z value it  slightly increases with the N/Z of the system. 
Assuming that the Coulomb
and thermal components of the kinetic energy are identical for all Xe+Sn
systems at a given incident energy, we induce that the expansion component
is larger, because heavier \emph {primary} isotopes are produced when more
neutrons are available in the system. The fragment kinetic energies obtained
in the SMF simulations are far below the experimental ones, due to the too
large preequilibrium emission (see section~\ref{sect63}).

%%%%%%%%%%%%%%%%%%%%%%%%%%%%%%%%%%%%%%%%%%%%%%%%%%%%%%%%%%%%%%%%%%%%%
\subsection{Summary \label{sect33}}
%%%%%%%%%%%%%%%%%%%%%%%%%%%%%%%%%%%%%%%%%%%%%%%%%%%%%%%%%%%%%%%%%%%%%
Comparing Xe+Sn quasi-fusion reactions for different 
pro\-jectile-target couples, we
have observed experimentally that more fragments, with larger kinetic
energies, are produced when the system is neutron-richer. The SMF model
reproduces these trends, before the cooling stage, meaning that we are
dealing with effects due to the interaction, more precisely to the symmetry
energy. The charge bound in fragments appears as an interesting isospin
sensitive variable, for which the simulation  better reproduces the trend 
(slope) with the asy-soft EOS.

%%%%%%%%%%%%%%%%%%%%%%%%%%%%%%%%%%%%%%%%%%%%%%%%%%%%%%%%%%%%%%%%%%%%%%%%%%%%
\section{Symmetry energy from isospin diffusion\label{sect4}}
%%%%%%%%%%%%%%%%%%%%%%%%%%%%%%%%%%%%%%%%%%%%%%%%%%%%%%%%%%%%%%%%%%%%%%%%%%%%
Isospin transport during semi-peripheral collisions between projectiles and
targets differing by their isospin content is one of the observables
sensitive to the symmetry term of the nuclear equation of state.
Around the Fermi energy the isospin content of the quasi-projectile (QP) and 
the quasi-target (QT) is determined by the interplay between fast particle 
emission during the overlap between the incident partners and the transfer 
of nucleons through the neck which develops between them~\cite{Bara05}. 
Assuming local thermal equilibrium the isospin transport coefficients can be
derived within the hydrodynamic limit; the proton and neutron exchange is
governed by the gradients of the respective chemical potentials and the
current of the two species comprises two terms. One is the isospin migration
arising from density gradient. The second one is the isospin diffusion due
to the different isospin content of the reaction partners~\cite{Bar05,Dit09}. 
The difference of the neutron and proton currents is directly connected to
the symmetry energy: the difference of the migration coefficients  is 
proportional to the isospin value times the slope of the symmetry energy 
versus density whereas that of the diffusion coefficients is proportional 
to the density times the absolute value of the symmetry energy.

It comes from the above considerations that even for a symmetric system
(identical projectile and target) one may observe some isospin content
evolution of the QP and QT with the incident energy and the impact
parameter, due to preequilibrium emission and isospin migration.

%%%%%%%%%%%%%%%%%%%%%%%%%%%%%%%%%%%%%%%%%%%%%%%%%%%%%%%%%%%%%%%%%%%%%%%%%%%%
\subsection{Isospin transport}\label{sect41}
%%%%%%%%%%%%%%%%%%%%%%%%%%%%%%%%%%%%%%%%%%%%%%%%%%%%%%%%%%%%%%%%%%%%%%%%%%%%
The INDRA collaboration studied isospin transport by looking at the
quasi-projectile isospin content, as both detection and isotopic
identification are better in the forward part of the INDRA array. Isospin
transport was studied as a function of the impact parameter (parameterized 
either \emph{via} the dissipated energy or \emph{via} the transverse 
energy of the light charged particles, see section~\ref{sect22}).
Information on the stiffness of  the symmetry energy was derived
through a comparison of the experimental data to the results of the
Stochastic Mean Field transport model briefly described in 
section~\ref{sect31}.

The INDRA collaboration studied isospin transport for different 
systems. The reader is sent to the published papers for details of the 
experiments; selections are described in section~\ref{sect22}.

%%%%%%%%%%%%%%%%%%%%%%%%%%%%%%%%%%%%%%%%%%%%%%%%%%%%%%%%%%%%%%%%%%%%%%%%
\subsubsection{Ni induced reactions \label{sect411}}
%%%%%%%%%%%%%%%%%%%%%%%%%%%%%%%%%%%%%%%%%%%%%%%%%%%%%%%%%%%%%%%%%%%%%%%%

In a first experiment a \nuc{58}{Ni} projectile accelerated at 52 
and 74~\AM{} by the GANIL facility
bombarded \nuc{58}{Ni} and \nuc{197}{Au}
targets~\cite{I71-Gal09,Gal09,I80-Gal10}.
The QP isospin variation was followed thanks to the variable
\begin{equation}
\mathrm{(<N>/<Z>)_{CP}} = \sum_{N_{evts}}{\sum_{\nu} {N_{\nu}}}
/ \sum_{N_{evts}}{\sum_{\nu} {P_{\nu}}}
\end{equation}
where $N_{\nu}$ and $P_{\nu}$ are respectively the numbers of neutrons and
protons bound in particle $\nu$ , $\nu$ being d, t, $^3$He, $^4$He, $^6$He,
$^6$Li, $^7$Li, $^8$Li, $^9$Li, $^7$Be, $^9$Be, $^{10}$Be; free protons are
excluded.  $N_{evts}$ is the number of events contained in the dissipated
energy bin considered. 
The variable $(N/Z)_{CP}$ was calculated twice: first considering
particles forward emitted in the nucleon-nucleon frame
($\mathrm{V_{particle}>V_{proj}^{lab}}/2$), and secondly keeping
only particles forward emitted in the QP frame 
($\mathrm{V_{particle}>V_{QP}^{rec}}$). In the first case mid-rapidity
particles and those coming from the hot QP de-excitation are considered. In
the second case, only the latter type of particles are kept.

%%%%%%%%%%%%%%%%%%%%%%%%%%%%%%%%%%%%%%%%%%%%%%%%%%%%%%%%%%%%%%%%%%%%%%%%%%%%%%
\begin{figure}[htb]
\includegraphics[width=\columnwidth]{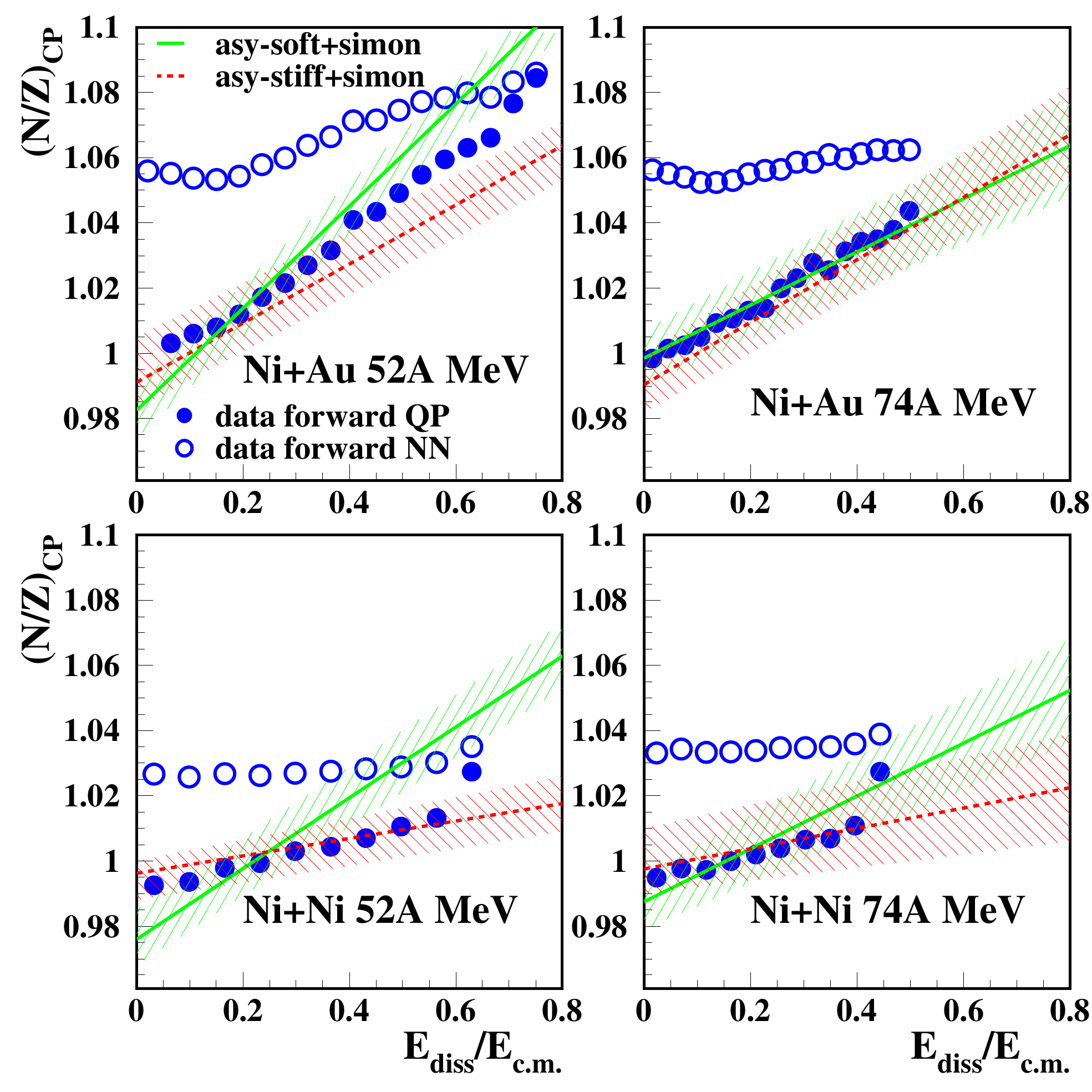}
\caption{(Colour on line) Isospin ratio of complex particles, $(N/Z)_{CP}$, 
vs the dissipated energy. Circles correspond to experimental data: open,
forward in NN frame, close, forward in QP frame. Error bars 
are within the size of the symbols. Dotted and solid lines correspond
respectively to the asystiff and asysoft parametrizations, the hatched 
zones give error bars from simulations. From.~\protect\cite{Gal09}}
\label{isodiff_Ni}
\end{figure}
%%%%%%%%%%%%%%%%%%%%%%%%%%%%%%%%%%%%%%%%%%%%%%%%%%%%%%%%%%%%%%%%%%%%%%%%%%%%%%%%
Figure~\ref{isodiff_Ni} displays the evolution of $(N/Z)_{CP}$ with the 
violence of the collision.
Open points show the values obtained forward 
in the NN frame. In this case we mix mid-rapidity
particles and those coming from the QP de-excitation.
For the Ni+Ni system at both incident energies, $(N/Z)_{CP}$
varies by at most 1.5\% when dissipation increases. This is the expected
behaviour for this symmetric system where N/Z is only modified by
pre-equilibrium emission.
For the Ni+Au system the isospin ratio is higher
than that of the Ni+Ni system whatever the dissipated energy.
At small dissipation this could arise from
the neutron skin of the Au target and/or from the mid-rapidity
particles included in our quasi-projectile selection, which are more neutron 
rich~\cite{I23-Lef00,I17-Pla99}. This result is a first indication of 
isospin diffusion.
Then $(N/Z)_{CP}$ presents a significant increase
with dissipation and reaches higher values at 52\AM, 
while the trend is flatter at 74\AM.
This may be interpreted as a progressive isospin diffusion when 
collisions become more central, in connection with a larger overlap 
of the reaction partners and thus a longer interaction time. 

The close points in fig~\ref{isodiff_Ni} are related to the values of
$(N/Z)_{CP}$ forward in the QP frame. They are in all cases smaller than the
previous ones, and for Ni+Au at both energies, they grow faster 
with dissipation. This is because the neutron-rich mid-rapidity particles 
are no  longer included; indeed it is known that their isospin content is 
independent of the violence of the collision.~\cite{I23-Lef00}

The values of $(N/Z)_{CP}$ forward in the QP frame are compared with the 
results of the SMF simulations, after de-excitation of the hot QP, 
displayed in the figure by the lines and the hatched zones.  
A first result  worth mentioning is that the chemical composition (N/Z) 
of the quasi-projectile forward emission appears as a very good 
representation of the composition of the entire quasi-projectile source. 
Such an observation seems to validate \emph{a posteriori} the
selection frequently used to characterize the QP de-excitation 
properties. When looking \emph{globally} at the results for the four
cases treated here, the agreement is better when the asy-stiff EOS is used,
i.e. a linear increase of the potential term of the symmetry energy around
normal density.  Note however that for Ni+Au at 52~\AM{}, where isospin
transport effects are dominant, the close points lie in between the 
simulated results  with the two EOS. This observation allows us to put an
error bar on our result, expressed as $\gamma$=1$\pm$0.2 
(see Eq.~\ref{Esym_gam}), or L=75$\pm$25~MeV (Eq.~\ref{Esym_L}).

%%%%%%%%%%%%%%%%%%%%%%%%%%%%%%%%%%%%%%%%%%%%%%%%%%%%%%%%%%%%%%%%%%%%%%%%
\subsubsection{Xe induced reactions \label{sect412}}
%%%%%%%%%%%%%%%%%%%%%%%%%%%%%%%%%%%%%%%%%%%%%%%%%%%%%%%%%%%%%%%%%%%%%%%%
%%%%%%%%%%%%%%%%%%%%%%%%%%%%%%%%%%%%%%%%%%%%%%%%%%%%%%%%%%%%%%%%%%%%%%%%
\begin{figure}[!bt]
\centering \includegraphics[width=\columnwidth]{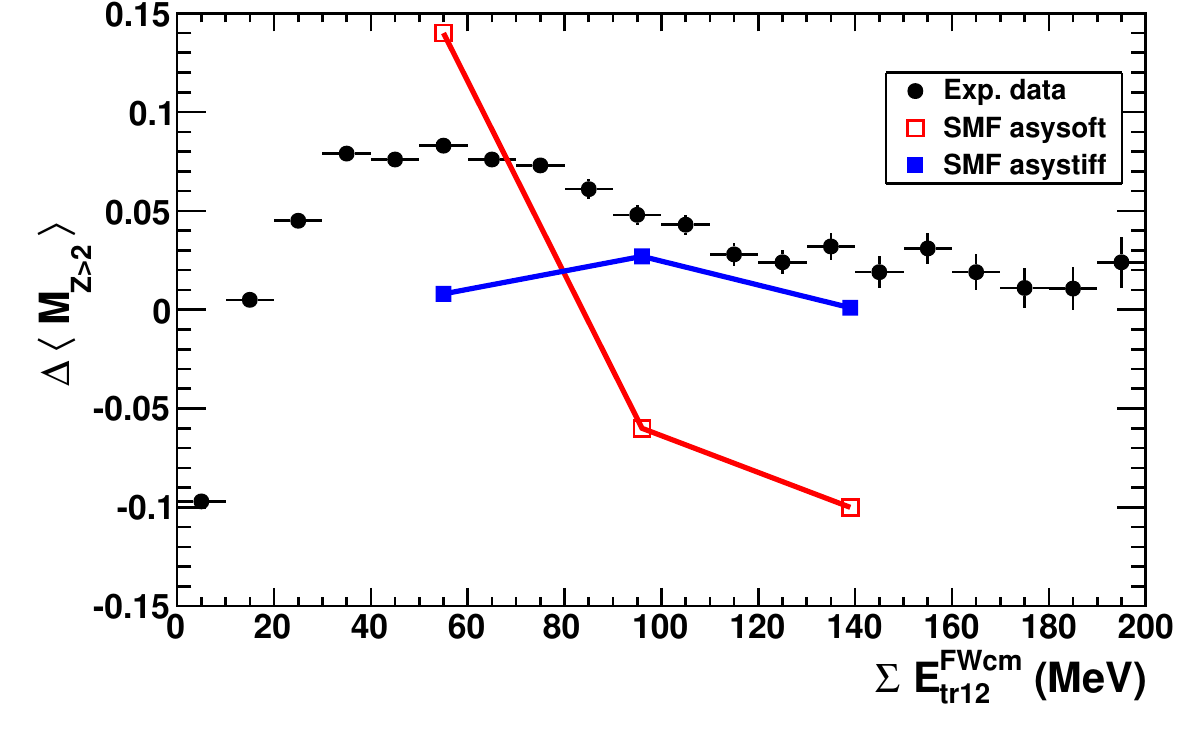}
\caption{(Colour on line) Difference of the fragment (Z$>$2) multiplicities, 
for the systems, 
\sys{136}{Xe}{112}{Sn} and \sys{124}{Xe}{124}{Sn}, \emph{vs} the transverse 
energy of light charged particles detected in the forward
cm hemisphere. Close points show the experimental data with statistical
error bars. Open squares correspond to the asysoft SMF simulation, 
close squares to the asystiff one. For simulations, the difference is 
calculated for primary fragments, before de-excitation. 
From.~\protect\cite{KabIWM11,T49Kab13}}
\label{isodiff_Xe}
\end{figure}
%%%%%%%%%%%%%%%%%%%%%%%%%%%%%%%%%%%%%%%%%%%%%%%%%%%%%%%%%%%%%%%%%%%%%%%%

A second study used 32~\AM{} \nuc{124,136}{Xe} projectiles impinging on
\nuc{112,124}{Sn}, as described in section~\ref{sect3}.  
It is interesting to note that the total systems obtained in the 
Ni+Au reaction at 52~\AM{}, and the \sys{124}{Xe}{124}{Sn} and 
\sys{136}{Xe}{112}{Sn} reactions at 32~\AM{}, are very similar and have 
the same N/Z. The available centre of mass energies
per nucleon differ by 1~\AM{}, being $\sim$9~MeV for Ni+Au at 52~\AM{} 
and $\sim$8~MeV Xe+Sn at 32~\AM{}. 

The chosen isospin sensitive variable is the fragment multiplicity difference 
between the \sys{136}{Xe}{112}{Sn} and \\ 
\sys{124}{Xe}{124}{Sn} systems: \\
\centerline{
$\Delta M_{Z>2} =  M_{Z>2}^{^{136}Xe+^{112}Sn} -
M_{Z>2}^{^{124}Xe+^{124}Sn}$} \\
followed as a function of the violence of the collision. 
The impact parameter scale is given by the transverse energy of the light 
charged particles: values $E_{tr12}^{FWcm}$ of 50, 100 and 140 MeV 
correspond to experimental impact parameters b=8, 6 and 4~fm respectively.
Only products detected in the forward centre of mass hemisphere are 
considered. Quasi-fusion events are
removed by means of the variable $V_{bigiso}$ (see section~\ref{sect21}). 

The measured $M_{Z>2}$ is the sum of
fragments coming from the QP de-excitation and from mid-rapidity. 
For $E_{tr12}^{FWcm} >$ 100 MeV  isospin equilibrium is
reached (see section~\ref{sect42}), thus the QP de-excitation properties 
are the same, in particular the multiplicities of the emitted fragments. 
Therefore $\Delta M_{Z>2}$ reduces to the
difference between mid-rapidity multiplicities. Assuming that these
multiplicities are not modified by the de-excitation stage, the measured 
$\Delta M_{Z>2}$ can be directly compared to the same difference obtained 
in SMF simulations for \emph{primary} fragments. This avoids to resort to 
a de-excitation code. Looking at fig.~\ref{isodiff_Xe}, one observes that
the experimental value of $\Delta M_{Z>2}$ levels-off above 
$E_{t12}^{FWcm}$=110~MeV. 
If we now compare with the simulated values, it appears that the asy-soft
case does not follow the experimental trend, whereas the asystiff
calculation well matches the data for $E_{tr12}^{FWcm} >$ 100 MeV, or
$b<6$~fm. For more peripheral collisions, the above assumptions do not hold
because isospin equilibrium is not reached, thus simulations and data diverge.

%%%%%%%%%%%%%%%%%%%%%%%%%%%%%%%%%%%%%%%%%%%%%%%%%%%%%%%%%%%%%%%%%%%%%%%%%%%%%
\subsection{Isospin equilibration}\label{sect42}
%%%%%%%%%%%%%%%%%%%%%%%%%%%%%%%%%%%%%%%%%%%%%%%%%%%%%%%%%%%%%%%%%%%%%%%%%%%%%

Ultimately isospin transport tends to equilibrate the isospin content of the
two reaction partners. Whether equilibrium is reached or not depends on the
relative values of the characteristic time of the mode, and of the reaction
time (namely when the QP and QT separate). In deep inelastic reactions at 
energies close to the Coulomb barrier, the N/Z equilibrium is rapidly reached, 
within $\sim$1-2$\times 10^{-22}$s~\cite{Gal76,Chi79}. In that case isospin 
equilibration is governed by a large amplitude dipole collective
motion~\cite{Dit09}. Around Fermi energy the experimental situation is not 
so clear. The
first experiments in the 80's studied Ar and Kr induced very peripheral 
reactions, in which isospin transport was visible but equilibrium was clearly 
not reached~\cite{Bor86}. In most of the recent papers, data  obtained at these 
energies, with some selection, were compared to transport calculations performed
at a given, supposedly matching impact parameter. At those energies, as said
above, the chemical potential gradient governs isospin exchange.  
%%%%%%%%%%%%%%%%%%%%%%%%%%%%%%%%%%%%%%%%%%%%%%%%%%%%%%%%%%%%%%%%%%%%%%%%%%%%%
\begin{figure}[!htb]
\centering \includegraphics[width=\columnwidth]{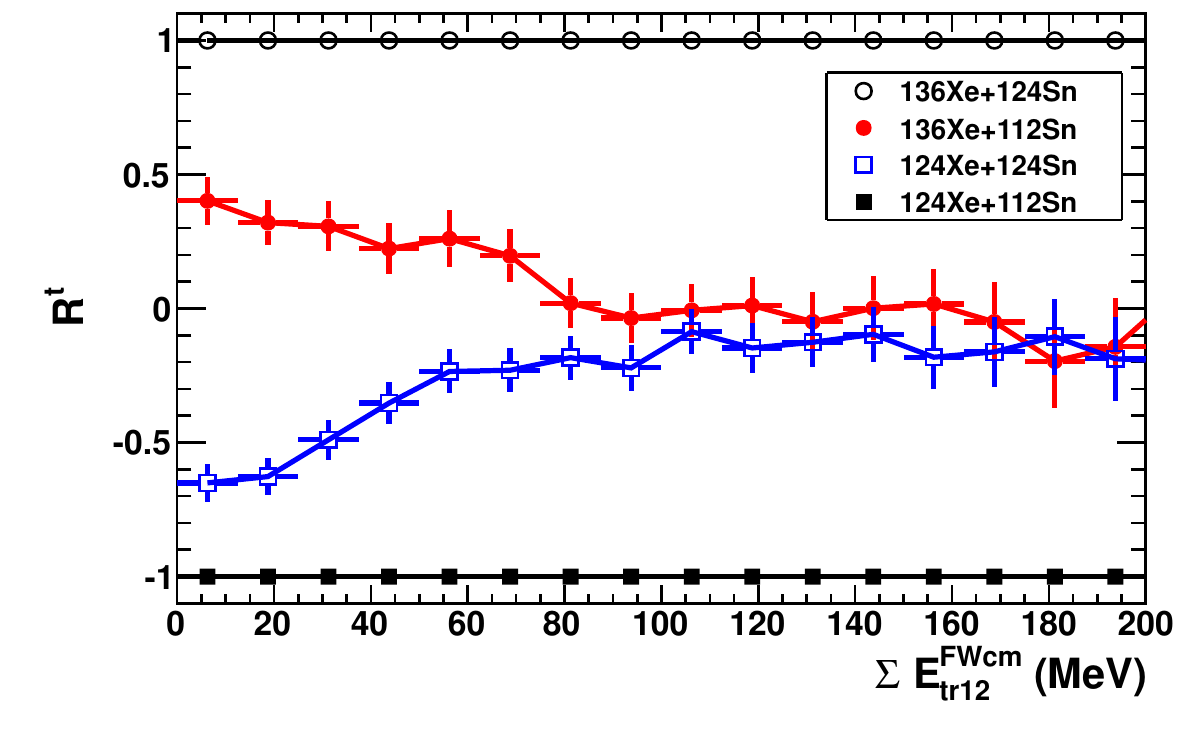}
\caption{Isospin transport ratio calculated for forward triton multiplicity
measured for the four Xe+Sn systems at 32~\AM{} as a function of the 
dissipated energy. From.~\protect\cite{T49Kab13}}
\label{Rt}
\end{figure}
%%%%%%%%%%%%%%%%%%%%%%%%%%%%%%%%%%%%%%%%%%%%%%%%%%%%%%%%%%%%%%%%%%%%%%%%%%%%%

As demonstrated in subsection~\ref{sect41}, we have the possibility 
to follow isospin 
transport from peripheral to almost central collisions, and to
anchor rather firmly the experimental and calculated impact parameters. 
We have arguments to state, from the experimental data alone, that in two
cases  isospin equilibrium is reached.
Looking at the top-left panel of fig.~\ref{isodiff_Ni}, we observe that at
high dissipation ($E_{diss}/E_{c.m.}>$0.7, i.e. $b<$5 fm from fig.~\ref{fig2})
the values of $(N/Z)_{CP}$ are the same, whether it is
calculated with or without mid-rapidity particles. This is a strong
indication of isospin equilibrium for the  \sys{58}{Ni}{197}{Au} reaction 
at 52~\AM{}. 
As far as the Xe+Sn systems are concerned, we show in fig.~\ref{Rt} the
isospin transport ratio~\cite{Riz08}

\begin{equation}
R^x_{P,T} = \frac{2(x^M-x^{eq})}{(x^H-x^L)} 
\ \mathrm{with} \ x^{eq}=(x^H+x^L)/2
\end{equation}
the index H refers to the n-rich system \sys{136}{Xe}{124}{Sn} and L to the n-poor 
\sys{124}{Xe}{112}{Sn} system, M to the mixed reactions \sys{124}{Xe}{124}{Sn}
and \sys{136}{Xe}{112}{Sn}. The chosen isospin sensitive variable is
the triton multiplicity in the forward centre of mass hemisphere. 
The evolution of $R^t$ with dissipation is displayed in fig.~\ref{Rt}.
We observe that isospin equilibrium is reached above a transverse LCP energy
of 100 MeV. The same conclusion is obtained if the variable x is the
fragment multiplicity~\cite{T49Kab13}.

Finally we studied almost symmetric and a very asymmetric reactions for which 
the composite systems have charges 104 and 107 and a value of N/Z of 1.38, at
available energies 8-9~\AM{}. We observed experimentally that isospin
equilibrium is reached for rather high dissipation.
To go further we turn to the SMF simulations. For Ni+Au at 52~\AM{}, isospin 
equilibrium is predicted for an impact parameter b$\leq$4~fm~\cite{Gal09}. 
The corresponding reaction time is 4$\times 10^{-22}$s, which puts an upper 
limit on the characteristic time for equilibrium.  
For the 32~\AM{} Xe+Sn systems SMF finds isospin equilibrium for impact
parameters smaller than 5-6~fm, corresponding to a reaction time of 
6-7$\times 10^{-22}$s. For b=8~fm, isospin equilibrium is not reached for a
reaction time of 5$\times 10^{-22}$s. Thus the isospin equilibration time
for Xe+Sn at 32~\AM{} can be estimated within 5 and 7$\times 10^{-22}$s.
The order of magnitude of this time is therefore similar for the two studied
systems at Fermi energies. It is around 4 times larger than the one obtained 
near the interaction barrier. 

%%%%%%%%%%%%%%%%%%%%%%%%%%%%%%%%%%%%%%%%%%%%%%%%%%%%%%%%%%%%%%%%%%%%%
\subsection{Conclusions on isospin transport} \label{sect43}
%%%%%%%%%%%%%%%%%%%%%%%%%%%%%%%%%%%%%%%%%%%%%%%%%%%%%%%%%%%%%%%%%%%%%
To summarize we have studied  isospin transport for several 
systems, using different sorting variables and isospin sensitive 
observables. We compared all data with the results of the
same stochastic mean field model. In all cases the agreement between
experimental and simulated results is better if the symmetry term of the
mean field is asy-stiff, with a potential part linearly increasing with 
 density. 
We also found that isospin equilibrium is reached for  available
energies around 8-9~\AM{}, independently of the mass asymmetry of the
entrance channel. This is also supported by the SMF simulations. 

If we compare with other published data, the preference for a potential
symmetry energy linearly increasing with density is also supported by time
evolution of the isospin of neck fragments~\cite{Def12}, compared with a 
slightly different SMF. And also by the results from the competition
between dissipative mechanisms for Ca induced reactions on Ca and Ti at 
25 \AM{}~\cite{Amo09}, compared with a CoMD simulation.
Results from Sn+Sn at 35 and 50~\AM{} presented in reference~\cite{Tsa09}
(several isospin dependent variables), compared to an improved quantum 
molecular dynamics model, ImQMD, plead for an asy-softer EOS. All these 
results fit however in the large limits defined in the
$L-S_0$ plane, see for instance figure~2 of ref.~\cite{Tsa12}.

Ref~\cite{Sun10} states that there is no isospin equilibrium for Sn+Sn at
35~\AM{}, in contradiction with our results for the close system
Xe+Sn at 32~\AM{}. Thus considering the state of the art on this subject, 
we have to recognize that extracting the asy-stiffness of the nuclear EOS 
from isospin diffusion is still an issue.

%%%%%%%%%%%%%%%%%%%%%%%%%%%%%%%%%%%%%%%%%%%%%%%%%%%%%%%%%%%%%%%%%%%%%%%%%%%%%%%%%%%%%%
\section{Isospin effects and energy dissipation \label{sect5}}
%%%%%%%%%%%%%%%%%%%%%%%%%%%%%%%%%%%%%%%%%%%%%%%%%%%%%%%%%%%%%%%%%%%%%%%%%%%%%%%%%%%%%%
Nuclear stopping observed in nuclear collisions can be used as a probe 
for transport properties of nuclear matter~\cite{I73-Leh10}. 
Studies of transport phenomena
are of primary importance for understanding the fundamental properties of 
nuclear matter such as energy dissipation, in-medium nucleon-nucleon cross 
sections ($\sigma_{nn/pp}$ and $\sigma_{np}$) and related mean free paths
or isospin diffusion~\cite[and refs. therein]{Dur00}. 
They are critical in the description
of the supernova collapse and the formation of a neutron star \cite{Lat04}.
Transport properties of nuclear matter are also one of the basic ingredients for
microscopic models \cite[and refs. therein]{And06,Fuc06}. 
In this section, we will present some
results concerning the degree of stopping achieved in the most central
collisions including, in particular, various projectile-target isospin
combinations for the Xe + Sn system.
We use the large dataset
provided by the INDRA collaboration concerning symmetric systems, with total
mass between $70$ and $400$, and incident energy covering the full range of the Fermi
energy domain between 10 and 100~\AM{}. 

%%%%%%%%%%%%%%%%%%%%%%%%%%%%%%%%%%%%%%%%%%%%%%%%%%%%%%%%%%%%%%%%%%%%%%%%%%%%%%%%%%%%%%
\subsection{Nuclear stopping observable \label{sect51}}
\begin{figure}[!tb]
\includegraphics[width=1.05\columnwidth]{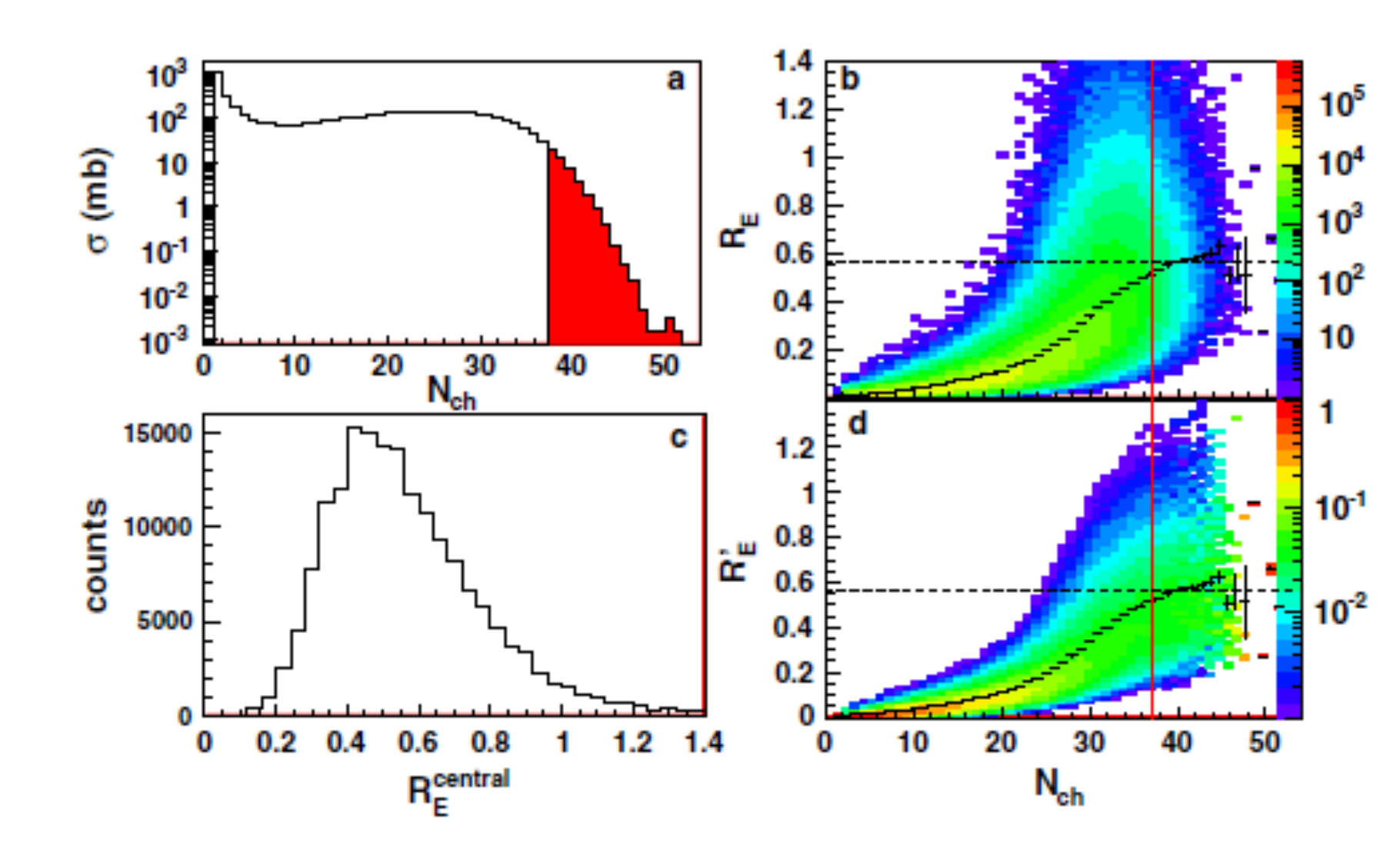}
\caption{(Colour on line) Total multiplicity for charged particles $N_{ch}$ (a), 
correlation between $R_E$ and $N_ch$ (b), distribution of $R_E$ corresponding 
to the red selection (c), and normalized correlation between $R_E$ and $N_{ch}$ (d).
From \cite{I73-Leh10}.}
\label{m_re}
\end{figure}

%%%%%%%%%%%%%%%%%%%%%%%%%%%%%%%%%%%%%%%%%%%%%%%%%%%%%%%%%%%%%%%%%%%%%%%%%%%%%%%%%%%%%%
Since we are here interested in the maximal values of stopping reached in 
nuclear collisions, we select the most central collisions, \emph{i.e.} 
collisions corresponding to the largest overlap between the two incoming nuclei.
The selection is here done using the total multiplicity of charged products $N_{ch}$
(see sect.~\ref{sect21}). 
To measure the degree of stopping we use the energy isotropy ratio $R_E$ defined on
an event-by-event basis as : 

\begin{equation}
R_E=\frac{1}{2} \frac{\sum_i E_i^{\perp}}{\sum_i E_i^{//}}
\end{equation}

where the index $i$ runs over the total number of charged particles in the event,
$E_i$ is the centre-of-mass kinetic energy for particle $i$, the indexes 
$\perp$ and $//$ stand
for transverse and parallel components of $E_i$ relative to the beam axis. By
construction, this ratio $R_E$ takes values between $0$ and $1$. 
Figure~\ref{m_re} displays (a) the total multiplicity 
distribution $N_{ch}$  and (b) the correlation between the isotropy
ratio $R_E$ and $N_{ch}$. 
%%%%%%%%%%%%%%%%%%%%%%%%%%%%%%%%%%%%%%%%%%%%%%%%%%%%%%%%%%%%%%%%%%%%%%%%%%%%%%
%%%%%%%%%%%%%%%%%%%%%%%%%%%%%%%%%%%%%%%%%%%%%%%%%%%%%%%%%%%%%%%%%%%%%%%%%%%%%%%%
We can see that $\langle R_E \rangle$ (depicted by the black lines in 
fig.~\ref{m_re}b and d) reaches an asymptotic value (here close to $0.6$) 
at large  $N_{ch}$ values. 
This allows to define a limit at  $N_{ch}=38$, indicated by the red line for the 
correlation in Fig.~\ref{m_re}b and the red histogram in the $N_{ch}$
distribution in Fig.~\ref{m_re}a, above which the isotropy ratio remains almost 
constant. The normalized correlation
between $R_E$ and $N_{ch}$ is also displayed onto Fig.~\ref{m_re}d.
The normalization is done on the \emph{z-axis}, by \emph{flattening} the 
$N_{ch}$ distribution to the same number of entries; this procedure allows to
compare  with the same level of statistics all values of $N_{ch}$ multiplicities.
We see on the normalized correlation in Fig.~\ref{m_re}d that the high values of 
$R_E$ ($R_E>1$) are significantly reduced, which could indicate that these large 
$R_E$ values are mainly due to the statistics. Fig.~\ref{m_re}c gives the 
resulting $R_E$ distribution given by the red selection on $N_{ch}$. From this 
distribution, we can extract the mean and standard deviation values of $R_E$. 
To conclude on this point, we have estimated the retained cross section by the 
total multiplicity selection; this corresponds to roughly $50-100$~mb, so to an 
impact parameter range between $0$ and $1-1.5$~fm, whatever
the system is.

%%%%%%%%%%%%%%%%%%%%%%%%%%%%%%%%%%%%%%%%%%%%%%%%%%%%%%%%%%%%%%%%%%%%%%%%%%%%%%%%
\subsection{Stopping for central collisions \label{sect52}}
%%%%%%%%%%%%%%%%%%%%%%%%%%%%%%%%%%%%%%%%%%%%%%%%%%%%%%%%%%%%%%%%%%%%%%%%%%%%%%%%
%%%%%%%%%%%%%%%%%%%%%%%%%%%%%%%%%%%%%%%%%%%%%%%%%%%%%%%%%%%%%%%%%%%%%%%%%%%%%%
\begin{figure}[!tb]
\includegraphics[width=1.1\columnwidth]{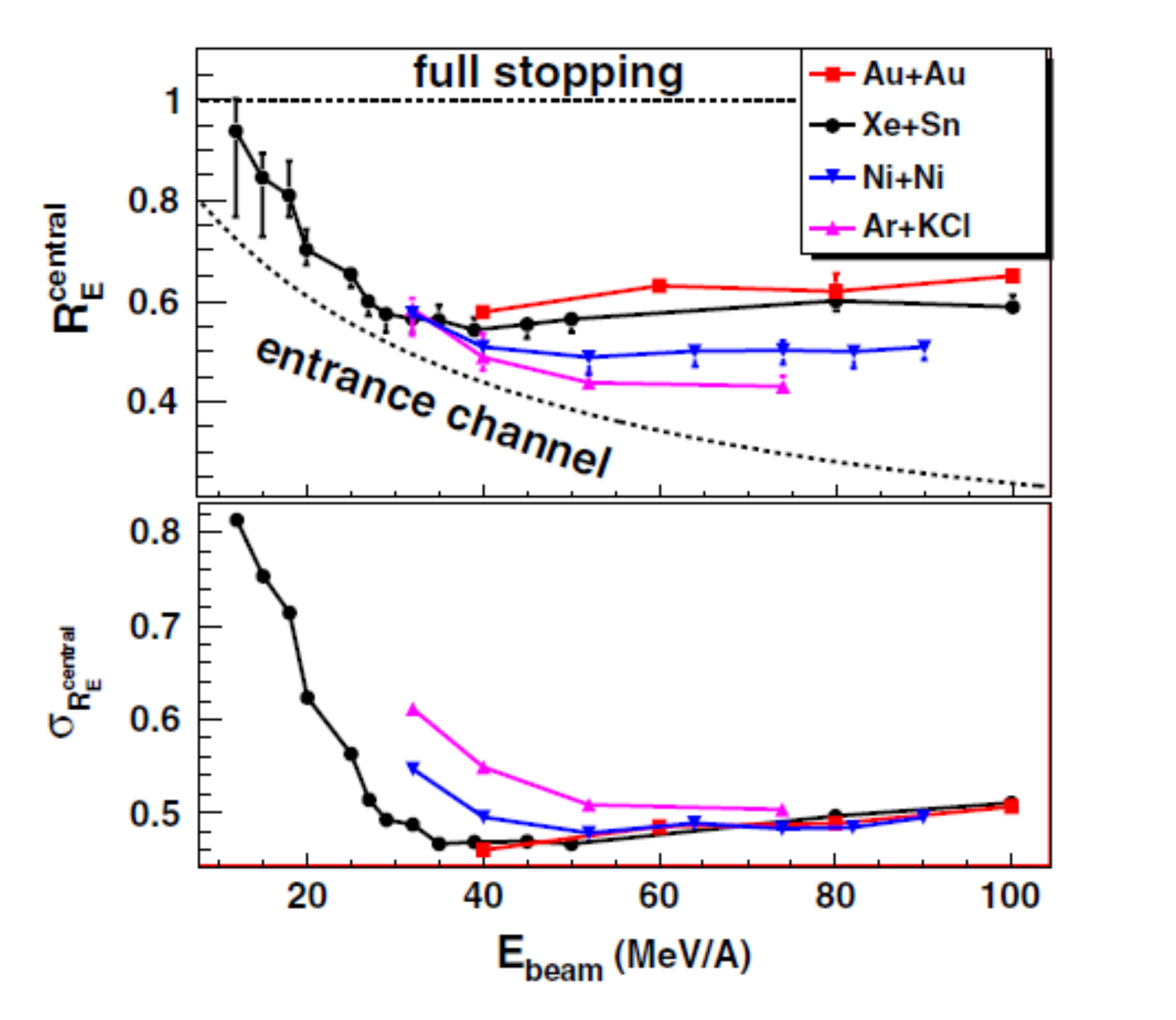}
\caption{(Colour on line) Isotropy ratio $R_E$ (top: mean values and bottom: 
rms) as a function 
of incident energy for various symmetric systems from central collisions.
From \cite{I73-Leh10}.}
\label{re_stopping}
\end{figure}
%%%%%%%%%%%%%%%%%%%%%%%%%%%%%%%%%%%%%%%%%%%%%%%%%%%%%%%%%%%%%%%%%%%%%%%%%%%%%%%%
%%%%%%%%%%%%%%%%%%%%%%%%%%%%%%%%%%%%%%%%%%%%%%%%%%%%%%%%%%%%%%%%%%%%%%%%%%%%%%%%
\begin{table*}[!t]
\centering
\begin{tabular}{| c | c | c | c | c |}
\hline
System & $\delta=\frac{(N_{tot}-Z_{tot})}{A_{tot}}$ & 32\AM{} & 45\AM{} & 100\AM{} \\
\hline
$^{124}Xe + ^{112}Sn$ & 0.119 & 0.54$\pm 0.04$ & 0.53$\pm 0.04$& 0.58$\pm 0.05$\\
$^{129}Xe + ^{112}Sn$ & 0.137 & - & - & 0.60$\pm 0.05$ \\
$^{124}Xe + ^{124}Sn$ & 0.161 & 0.54$\pm 0.04$& - & 0.56$\pm 0.04$\\
$^{129}Xe + ^{nat}Sn$ & 0.161 & 0.55$\pm 0.03$ & 0.53$\pm 0.04$ & - \\
$^{136}Xe + ^{112}Sn$ & 0.161 & 0.50$\pm 0.03$ & 0.54$\pm 0.04$ & - \\
$^{129}Xe + ^{124}Sn$ & 0.178 & - & - & 0.59 $\pm 0.05$\\
$^{136}Xe + ^{124}Sn$ & 0.200 & 0.49$\pm 0.03$ & 0.52$\pm 0.05$ & -\\
\hline
\end{tabular}
\caption{Mean values and standard deviations for $R_E$ in central collisions.}
\label{Re_isospin}
\end{table*}
%%%%%%%%%%%%%%%%%%%%%%%%%%%%%%%%%%%%%%%%%%%%%%%%%%%%%%%%%%%%%%%%%%%%%%%%%%%%%%%%%
We have applied the same procedure for all symmetric systems available in
the INDRA collaboration datasets. The mean and standard deviation values of
the selected $R_E$ distributions  are presented in 
Fig.~\ref{re_stopping}. 
We notice the evolution of $R_E$, from almost $1$ for low incident energy to 
values in between the two limits represented by the dashed curves labeled 
"entrance channel" and "full stopping". They indicate the $R_E$ values for a 
full stopping ($R_E=1$) and for no stopping at all; in this latter case, 
we can simply estimate the expectation values from a simulation of two Fermi 
spheres in momentum space separated by the relative momentum corresponding to the
entrance channel. Hence, this lower limit represents the minimal value of $R_E$ 
in absence of any stopping. We clearly observe two regimes for the stopping;
the first one corresponds to low incident energy up to 30-35~\AM{}. 
The decreasing stopping  is attributed to the progressive
disappearance of the nuclear (Mean-Field) effects ruled by 1-Body
dissipation~\cite{I73-Leh10}. The second regime corresponds to the high incident 
energy range, from the Fermi energy up to the highest available energy
$E_{inc}$=100~\AM{}.
The isotropy ratio remains rather flat, with a clear mass hierarchy : the larger 
the system, the higher $R_E$ is. This phenomenon is attributed to the appareance
of the nucleonic degrees of freedom and 2-Body dissipation induced by (elastic)
nucleon-nucleon collisions \cite{I73-Leh10}. Indeed, these latter become more and
more likely as we move away from the Fermi energy and, in a \emph{Glauber} picture, 
become more and more abundant as the number of participants increases.

%%%%%%%%%%%%%%%%%%%%%%%%%%%%%%%%%%%%%%%%%%%%%%%%%%%%%%%%%%%%%%%%%%%%%%%%%%%%%%%%%
\subsection{Isospin effect on nuclear stopping \label{sect53}}
%%%%%%%%%%%%%%%%%%%%%%%%%%%%%%%%%%%%%%%%%%%%%%%%%%%%%%%%%%%%%%%%%%%%%%%%%%%%%%%%%
We have seen previously that the isotropy ratio can be used as a probe to 
the energy dissipation/stopping achieved in central collisions. In this section, 
we give a special emphasis of the eventual isospin effect on the isotropy ratio. 
To do so, we use various isospin combinations of INDRA Xe+Sn systems: 
$^{124,129,136}$Xe, $^{112,nat=119,124}$Sn at different incident energies: 
32, 45 and 100~\AM{}. The same event selection (central collisions selected 
from $N_{ch}$) has been used. We display the results for the corresponding 
isotropy ratio in table \ref{Re_isospin}.

We show that, except for the \nuc{136}{Xe} induced collisions at $E_{inc}/A=$ 
32~MeV, all values  are statistically compatible at a given incident energy. 
This seems to indicate that the isospin content does not influence the isotropy 
ratio and hence the energy dissipation in the Fermi energy range up to a 
$\delta=(N-Z)/A$ close to $20\%$. This result is quite surprising since we expect 
some variations due to the differences on the magnitude of the
isovector channels for the nucleon-nucleon cross section $\sigma_{nn/pp}$ and
$\sigma_{np}$~\cite{Met58,Kik68,Li93}. 

%%%%%%%%%%%%%%%%%%%%%%%%%%%%%%%%%%%%%%%%%%%%%%%%%%%%%%%%%%%%%%%%%%%%%%%%%%%%%%%%%%%%%
\subsection{Conclusions \label{sect54}}
%%%%%%%%%%%%%%%%%%%%%%%%%%%%%%%%%%%%%%%%%%%%%%%%%%%%%%%%%%%%%%%%%%%%%%%%%%%%%%%%%%%%%
To conclude on these aspects related to the transport properties of nuclear 
matter, we believe that stopping studies in central collisions can provide 
new information about the fundamental \emph{in-medium} quantities such as the 
nucleon-nucleon mean free path, the nucleon-nucleon cross section for 
example~\cite{Che13,*Sam13}. Moreover, isovector
properties of the nuclear interaction like the (reduced) mass splitting between
protons and neutrons in the medium can be probed by using asymmetric nuclear 
matter systems produced with next-generation radioactive beam facilities and 
detection arrays. This could help to assess from a microscopical point of view 
the link between the isovector properties of the nuclear interaction and the 
potential part of the symmetry energy in the nuclear equation of state. As a 
perspective, all these results concerning the study of stopping in nuclear 
collisions speak in favour of more detailed analyses both on experimental and 
theoretical side.

%%%%%%%%%%%%%%%%%%%%%%%%%%%%%%%%%%%%%%%%%%%%%%%%%%%%%%%%%%%%%%%%%%%%%%%%%%%%%%%%%%%%%
\section{Studies in progress \label{sect6}}
%%%%%%%%%%%%%%%%%%%%%%%%%%%%%%%%%%%%%%%%%%%%%%%%%%%%%%%%%%%%%%%%%%%%%%%%%%%%%%%%%%%%%
\subsection{Symmetry energy, level density parameter and fission barriers
\label{sect61}}
%%%%%%%%%%%%%%%%%%%%%%%%%%%%%%%%%%%%%%%%%%%%%%%%%%%%%%%%%%%%%%%%%%%%%%%%%%%%%%%%%%%%
Technical advances both in producing exotic beams and in detection devices
allow to progress in the study of the de-excitation of hot nuclei formed by
heavy-ion induced fusion reactions. SPIRAL at GANIL now accelerates exotic
Ar beams up to 15~\AM{} and Kr beams around 5-6~\AM{}. 
The 4$\pi$ array INDRA has proven to be an efficient tool to measure 
simultaneously light charged particles, fission fragments and evaporation 
residues at low bombarding energy, and was used to scan the de-excitation of
medium-mass hot nuclei. Its drawback is the lack of mass
identification for the evaporation residues. We thus realized a very
ambitious experiment aiming at completely identifying the residue and all
emitted particles by coupling INDRA with the large acceptance VAMOS
spectrometer~\cite{Sav03,Pul08,Pull08}.

Understanding the de-excitation of hot nuclei requires the knowledge of
fundamental nuclear parameters, such as level density parameters and 
fission
barriers, to describe the thermal and collective properties that rule the
competition between the different decay modes. The isospin content of the 
compound nucleus has a
strong influence on these quantities. The fission barriers strongly depend 
on the symmetry energy, weakly constrained by the experimental 
data~\cite{Sie85}. 
The level density parameter, $a$, is related
to the effective mass, a property of the nuclear interaction that is
sensitive to the neutron and proton content of the nuclei. 
The effective nucleon mass is expected to decrease with increasing T while
T$\leq$2 MeV. This implies a decrease of the level density parameter but
also an increase with T of the kinetic symmetry energy contribution to the
nuclear binding energy $E_{sym} (T) = b_{sym} (T) \times (N-Z)^2/A$. These
effects would experimentally appear as a change in the particle multiplicity
and in the relative yields of the exit channels~\cite{BroCNR09}.
Experimentally $a$ can not be directly measured at high energy, but the 
temperature $T$ and  $1/T = \mathrm{d} \ln \rho / \mathrm{d} E^* $
can be extracted from the exponential slope of kinetic energy
spectra of evaporated particles. Multichance emission could be taken into 
account through statistical model calculations like GEMINI~\cite{Char88}. 
Comparisons with calculations~\cite{Shl91}  constrain the dependence of $a$ with
$E^*$ and $T$ and verify the consistency with other data for known isotopes.
$a$ was shown to evolve from A/8.5 MeV$^{-1}$ at low temperature to 
A/15 MeV$^{-1}$ around T=4-5 MeV~\cite{Neb86}.
The predicted isospin dependence of level density within the Fermi 
gas model is a small decrease with increasing (N-Z).
A significantly larger dependence would have important implications for
other fields (r-process for instance).  Different extrapolations starting from 
stable nuclei have been proposed~\cite{AlQ01} that lead to quite important 
variations on the estimated values of the level density parameter. 
It may even vanish when approaching
the proton-drip line. Experimental data far from the valley 
of stability are very scarce. For that reason, and also because in the
multifragmentation process observed at Fermi energies, excited neutron
deficient fragments are assumed to be formed, and their de-excitation is not 
well constrained~\cite{Bor08}, it is important to get new information on the
de-excitation over a large range of isotopes of compound nuclei.

%%%%%%%%%%%%%%%%%%%%%%%%%%%%%%%%%%%%%%%%%%%%%%%%%%%%%%%%%%%%%%%%%%%%%%%%%%%%
\subsubsection{De-excitation of Ba nuclei formed at low bombarding energy}
\label{sect611}
%%%%%%%%%%%%%%%%%%%%%%%%%%%%%%%%%%%%%%%%%%%%%%%%%%%%%%%%%%%%%%%%%%%%%%%%%%%
Fission barriers strongly depend on the symmetry energy parameter used 
for calculating the macroscopic part of binding energies. This term is 
insufficiently constrained by experimental data on the fission 
barriers~\cite{Sie85} of nuclei with A=100--180, 
where nuclei formed in this intermediate-mass region are able to sustain
extreme stresses, including high temperatures, large rotation and
deformation. 

In this line an experiment was performed with INDRA to study 
the $^{78,82}$Kr+$^{40}$Ca reactions at 5.5~\AM{}~\cite{I76-Ade11}. 
At this low incident energy, around the Coulomb barrier,  
medium-mass compound nuclei are formed in a controlled way in terms of 
excitation energy and angular momentum.

The cross section, kinetic energy distribution, angular distribution
of fragments with atomic number  3 $\le Z  \le$ 28 and coincidences 
between light charged particles and fragments were measured.
Global features indicate a high degree of relaxation and  are 
compatible with a binary fission from compound nuclei.
%%%%%%%%%%%%%%%%%%%%%%%%%%%%%%%%%%%%%%%%%%%%%%%%%%%%%%%%%%%%%%%%%%%%%%%%%%%%%%%
\begin{figure}[h!]
  \centering
\includegraphics[width=0.8\columnwidth]{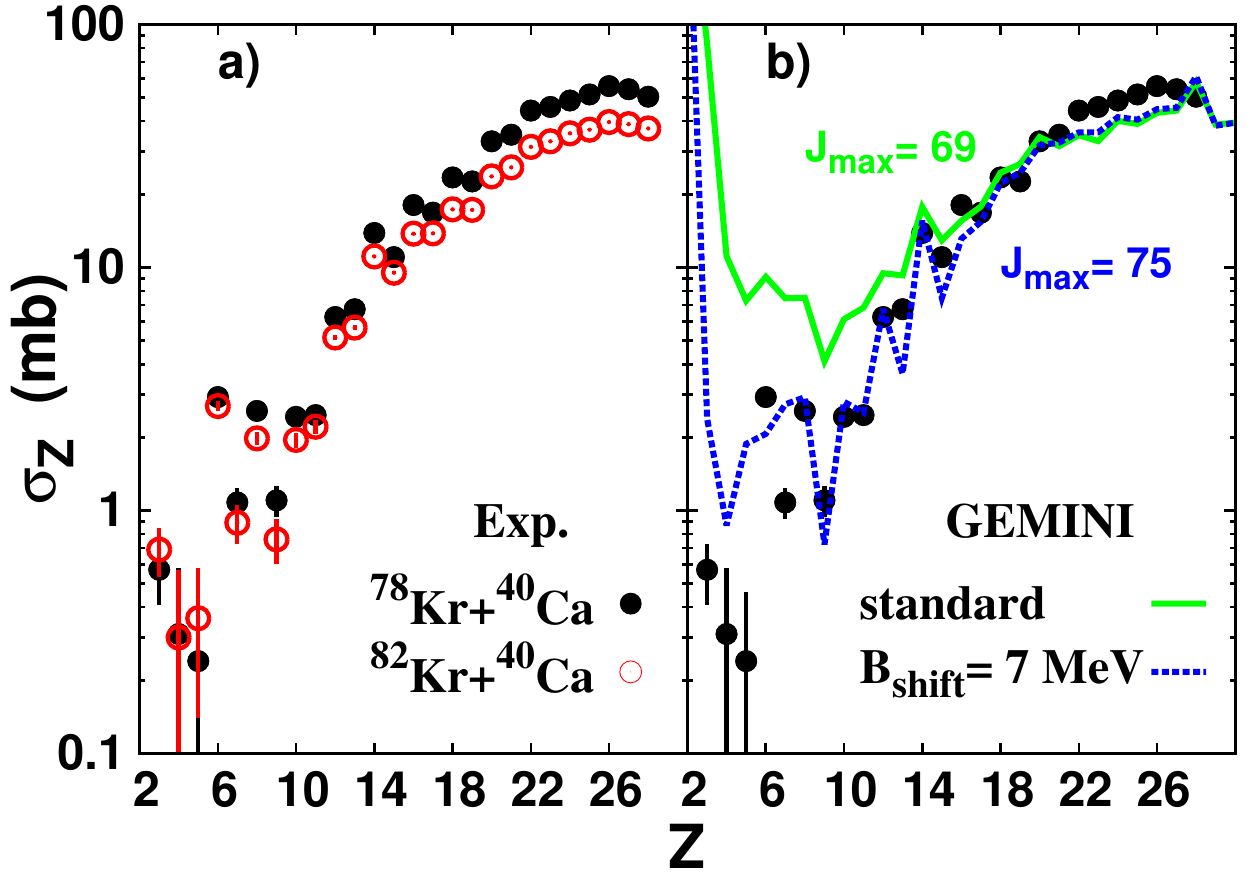}
\caption{(Colour on line) Experimental cross-sections for fragments 
measured in the
$^{78}$Kr+$^{40}$Ca (full circles) and  $^{82}$Kr+$^{40}$Ca (open circles) 
reactions at 5.5~\AM{}. Curves are the predictions of two calculations 
performed in the framework of the transition state model~\cite{Chari88} 
assuming the fission barriers given by the FRLDM~\cite{Sie85} (solid line) 
and by the FRLDM barriers increased by a constant B$_{shift}=7$~MeV 
(dashed line). A level density parameter $a=A/7$~MeV$^{-1}$ is taken 
for both calculations.}
\label{fig:9}       % Give a unique label
 \end{figure}
%%%%%%%%%%%%%%%%%%%%%%%%%%%%%%%%%%%%%%%%%%%%%%%%%%%%%%%%%%%%%%%%%%%%%%%%%%%%%%
Inclusive cross-section distributions of fragments with charge 
3 $\le Z  \le$ 28, displayed in fig.~\ref{fig:9}a, are typical of
evaporation plus fission distributions, and 
a strong odd-even staggering (oes) is observed for 3 $\le Z \le$ 12, 
indicating that structure effects persist. No isospin dependence
of the oes magnitude is clearly visible. The fission channel
is 25\% higher for the system with the lowest neutron-to-proton ratio 
while the evaporation residue cross-sections for both systems are similar 
within the error bars.

Coincidence  measurements  between light charged particles and fragments 
suggest that  the  light  partners in  very
asymmetric  fission  are  emitted  at excitation  energies  below  the
particle emission thresholds for both systems: the  odd-even staggering
of  the light fragments cross-sections   is not predominantly related
to the secondary  emission of light particles from excited fragments.
The $\sigma_{Z}$ of the light  fragments reflects directly the
primary fragmentation and  thus they  provide  important  constraints
on  the  energetic balance between both fragments.

The $Z$-dependence of the  mean value of the total kinetic energy (TKE)
indicates dominance of the Coulomb interaction between partners. 
The width of the TKE distributions signals large fluctuations of the TKE
and the ratio  $\sigma(TKE)$/$\langle TKE \rangle$ is roughly constant. 
The global features of the kinetic energy distributions are not 
reproduced assuming a total kinetic energy including both contribution
extracted from the Viola systematics~\cite{Vio85} and from  the relative motion.

Statistical calculations assuming  spherical fission fragments and  
finite-range liquid drop fission barriers are not able to explain the 
experimental features. A fitting procedure assuming a $TKE$ fluctuation 
given by an average value of the experimental one and a constant shift 
of 7~MeV on the Finite Range Liquid Drop Model (FRLDM) fission 
barriers~\cite{Sie85} allows to explain both  
the $Z$-distribution over the whole range of charge-asymmetry and 
the $Z$-dependence of the $\langle TKE \rangle$ except for  $Z\ge$20, see
fig.~\ref{fig:9}b. In a recent 
work~\cite{ManIWM11}, the increase of the fission barriers by such 
a shift was also proposed to reproduce charge distributions in similar
reactions at lower angular momenta.

The $TKE$ fluctuation could be related to deformations of the partners 
indicating the strong influence of the shape parameterization of the 
potential energy surface in describing the fission process of intermediate
mass compound nuclei.
%%%%%%%%%%%%%%%%%%%%%%%%%%%%%%%%%%%%%%%%%%%%%%%%%%%%%%%%%%%%%%%%%%%%%%%%%%%%
\subsubsection{Level density parameter of Pd nuclei from stability to the
proton-drip line} \label{sect612}
%%%%%%%%%%%%%%%%%%%%%%%%%%%%%%%%%%%%%%%%%%%%%%%%%%%%%%%%%%%%%%%%%%%%%%%%%%%%%
We  studied the de-excitation properties of Pd nuclei formed in 
collisions between different Ar projectiles and Ni targets:
\sys{34}{Ar}{58}{Ni}, \sys{36}{Ar}{58}{Ni}, \sys{36}{Ar}{60}{Ni},
\sys{40}{Ar}{60}{Ni} and \sys{40}{Ar}{64}{Ni},
at incident energies around 13 A.MeV, using VAMOS coupled with INDRA. 
This energy was a compromise between not too large 
preequilibrium effects and sufficient recoil energy for nuclear charge 
identification of residues. The exact incident energy for each beam was
chosen to get the same excitation energy per
nucleon of compound nuclei (2.9 MeV) with very similar angular momentum ranges.

The unstable \nuc{34}{Ar} beam was extremely important since it allowed to 
touch the  p-drip line in forming \nuc{92}{Pd}:
depending on model the drip-line is predicted to be between masses 84 and
89~\cite{Lal01}. In this case special de-excitation properties might be 
observed. 
With the stable \nuc{36}{Ar} beam coupled to the \nuc{60}{Ni} target 
the neutron-magic nucleus \nuc{96}{Pd} is made.

The detection of complete events should put an  additional strong constraint 
on the values of $a$ for nuclei along the de-excitation chain, provided by 
the correct weighing of the different exit channels; this was never measured 
up to now. As a simple example, in the 
experiment the Ni(Ar,$\alpha$xn)Ru channel can be distinguished from the 
Ni(Ar,2p(x+2)n)Ru and Ni(Ar,pd(x+1)n)Ru channels and correctly weighed.
All decay chains can be precisely characterized (isotopic composition of 
emitted particles and their multiplicity as well as the residue 
characteristics (A, Z) and their kinetic energies event by event) and 
we will obtain the percentage with which different chains lead to the same 
residue. Note that if the total detected charge is that of Pd, the neutron
multiplicity is simply derived by the difference between the compound
nucleus mass and those of all detected de-excitation charged products.

 Some preliminary data were published, which made use of the INDRA response
alone, without identification or calibration~\cite{T46Mar09,MarCNR09}. The
fusion-evaporation cross-section seems to decrease by a factor $\sim$2 for
the system close to the drip-line whereas it is constant for the four other
systems. INDRA is now fully calibrated, and the complete reconstruction of
residue properties in VAMOS is almost achieved. 
Concerning cross sections we will firstly  refine this preliminary analysis, 
with the full residue angular distributions (no measurements below 7$^o$ were 
performed with INDRA alone) and secondly determine the process
responsible of the missing cross-section. For similar systems (far from the
drip-line) fusion-evaporation and fusion-fission cross-sections were found
nearly equal~\cite{Gau75,Cof84}. A possibility is thus an increase of the 
fission cross-section due to a larger fissility parameter $Z^2/A$.
But as stated above, we might expect exotic processes when nuclei very close 
to the drip-line are formed. Moreover the compound nuclei excitation energies 
are close to the multifragmentation threshold. In a Lattice Gas framework, it
was shown that both isospin and Coulomb effects lower the transition temperature 
of the liquid-gas phase transition~\cite{Leh10}. 
Statistical model simulations (GEMINI) will be used to determine the values
of $a$, at least for the compound nuclei, including the strong constraints
given by the de-excitation processes cross-sections and the weighted
de-excitation chains.

%%%%%%%%%%%%%%%%%%%%%%%%%%%%%%%%%%%%%%%%%%%%%%%%%%%%%%%%%%%%%%%%%%%%%%%%%%%%%%%%%%%%%
\subsection{Symmetry energy and isotopic distributions \label{sect62}}
%%%%%%%%%%%%%%%%%%%%%%%%%%%%%%%%%%%%%%%%%%%%%%%%%%%%%%%%%%%%%%%%%%%%%%%%%%%%%%%%%%%%%
\begin{figure*}
\centering \includegraphics[width=0.75\textwidth]{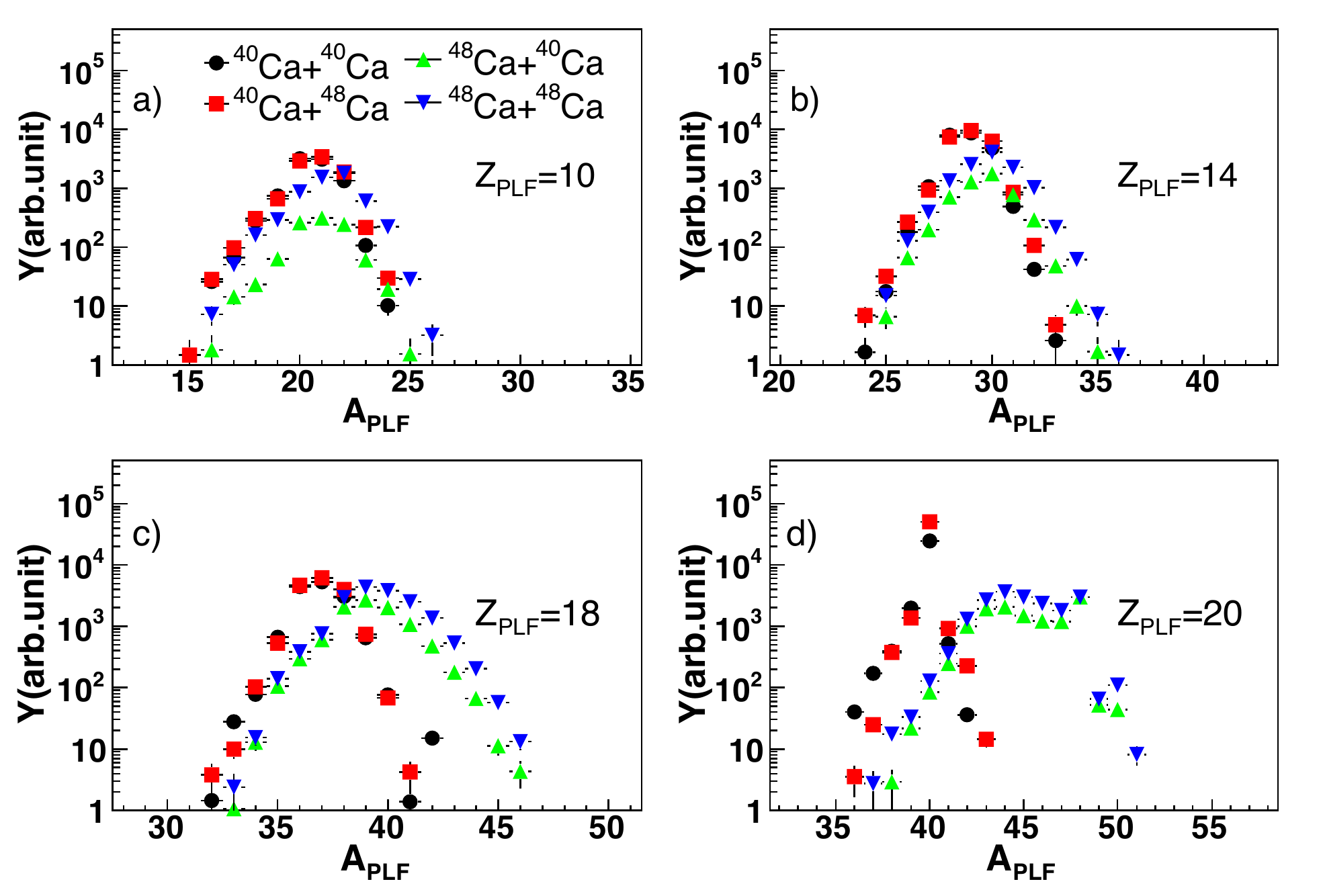}
\caption{(Colour on line) Isotopic distributions of fragments with atomic
number $Z_{PLF}$ = 10, 14, 18 and 20, produced in the reactions 
\sys{40,48}{Ca}{40,48}{Ca} at $E/A=$ 35 MeV.}
\label{fig:10}
\end{figure*}
%%%%%%%%%%%%%%%%%%%%%%%%%%%%%%%%%%%%%%%%%%%%%%%%%%%%%%%%%%%%%%%%%%%%%%%%%%%%%%%
Another observable predicted to be sensitive to the symmetry energy is
the shape of isotopic distributions. Such distributions were studied in
transport code simulations, AMD (Antisymmetrized Molecular 
Dynamics)~\cite{Ono99} and SMF~\cite{Colon98}, through
the determination of the free energy of fragments in the model 
connected with their statistical properties, like isoscaling (see
section \ref{sect72}).

Recently, in the framework of  Stochastic Mean Field  calculations, 
M.~Colonna~\cite{Col13} demonstrated that the shape of the isotopic 
distribution can be related to the symmetry energy term of the EOS. 
In this context, full SMF simulations in a box for unstable matter allow the 
fragment formation.  The 
isovector fluctuations were estimated as a function of the local density inside 
the fragmenting system by looking at the variance of the isovector density in 
cells having the same density. If the equilibrium is reached the quantity 
$F=T/\sigma$ (where $T$ is the temperature and  $\sigma$ is the isovector 
variance) coincides with the effective symmetry free energy. This analysis 
of the isotopic distribution of the fragments should thus probe the local
symmetry energy of clusterized systems (which is different from the
extraction of the total symmetry energy associated with clusterized
low-density matter~\cite[and refs. therein]{Col13}).

In ref.~\cite[and refs. therein]{Ono04} the authors simulate head-on 
nuclear collisions at 35~\AM{}; they construct a global isotopic distribution, 
K(N,Z), by combining together all the yields of fragments (N,Z) obtained in 
four reaction systems: 
\sys{40}{Ca}{40}{Ca}, \sys{48}{Ca}{48}{Ca},
\sys{60}{Ca}{60}{Ca}, and \sys{46}{Fe}{46}{Fe}, in order to get a broad
range of isotopes. 
The K(N,Z) 
distributions for each Z-value, were well fitted by a quadratic function
\begin{equation}
K(N,Z)=\eta(Z)+\xi(Z)N+\zeta(Z)\frac{(N-Z)^2}{N+Z}
\label{eqk}
\end{equation}
where $\eta(Z)$, $\xi(Z)$ and $\zeta(Z)$   are the fitting parameters. The obtained 
parameter   of the quadratic term in (N-Z) is associated by the authors to the 
symmetry energy coefficient $c_{sym}(A)$ in the EOS through the relation 
$\zeta(Z)=c_{sym}(A)/T$ , 
where T is the temperature of the system. The symmetry coefficient $c_{sym}(A)$ 
is here the sum of a volume and a surface term, 
$c_{sym}(A) = c_{sym}^v+c_{sym}^s A^{-1/3}$, as
in advanced mass formulae~\cite{Mye66,*Mol95,*Dan03}.
The obtained values of   $\zeta(Z)$ 
for each Z  have almost no dependence on the charge Z (Z $> $4) (see Fig.3 of 
Ref.~\cite{Ono04}).  The AMD predictions of constant  $\zeta(Z)$ values with 
increasing Z  indicate that the contribution of the surface term to the symmetry 
energy is strongly reduced in multifragmentation events. Based on these findings, 
the authors of Ref.~\cite{Ono04} conclude that, at the low density freeze-out stage,
the surface term does not contribute strongly to the symmetry energy. Therefore, 
the symmetry energy, $c_{sym}(A)$, at finite temperature and subsaturation 
densities that one can extract from fragment isotopic distributions corresponds 
to the volume term of the symmetry energy in infinite nuclear matter.
 This result requires experimental confirmation.

%%%%%%%%%%%%%%%%%%%%%%%%%%%%%%%%%%%%%%%%%%%%%%%%%%%%%%%%%%%%%%%%%%%%%%%
\subsubsection{Experimental methods}\label{sect621}
%%%%%%%%%%%%%%%%%%%%%%%%%%%%%%%%%%%%%%%%%%%%%%%%%%%%%%%%%%%%%%%%%%%%%%%

In view of the 
above considerations, we performed an experiment at GANIL using the VAMOS 
spectrometer~\cite{Sav03,Pul08,Pull08} coupled to the INDRA array. 
We studied four reactions \sys{40}{Ca}{40}{Ca}, \sys{40}{Ca}{48}{Ca},
\sys{48}{Ca}{40}{Ca} and \sys{48}{Ca}{48}{Ca} at E/A=35~MeV.
The use of 
a magnetic spectrometer allows one to measure a very wide range of isotopic 
distributions with the highest mass resolution. Coupling VAMOS and INDRA 
permits a good impact parameter determination and the use of calorimetry 
techniques to estimate excitation energies and temperatures in these reactions. 
In the case of peripheral collisions we measure with VAMOS the isotope production 
for PLF's produced by deep inelastic mechanism at low excitation energy (i.e.
with velocity close to that of the beam). The extraction 
of the  $\zeta$ parameter for each element can provide information about the 
surface effects in the symmetry energy at finite temperature and at densities 
close to the saturation density. The acceptance of the VAMOS spectrometer is
large enough to measure also isotopic distributions produced in more central 
collisions. In these more violent events, 
higher excitation energies are involved and we expect to approach the 
multifragmentation regime where a study of the isotopic distributions provides 
access to  $\zeta$ and the symmetry energy at sub-saturation densities.

%%%%%%%%%%%%%%%%%%%%%%%%%%%%%%%%%%%%%%%%%%%%%%%%%%%%%%%%%%%%%%%%%%%%%%%
\subsubsection{Isotopic distributions}\label{sect622}
%%%%%%%%%%%%%%%%%%%%%%%%%%%%%%%%%%%%%%%%%%%%%%%%%%%%%%%%%%%%%%%%%%%%%%%
Experimental isotopic distributions of fragments with $Z = 10, 14, 18$ 
and $20$ are presented in Fig.~\ref{fig:10} for the four analyzed systems. 
The distributions span over more than 10 isotopes with relative yields 
covering 4 orders of magnitude. 
Very n-rich isotopes are populated reaching a value of $N/Z=$ 1.6
for $Z_{PLF}=$10 and $N/Z=$ 1.57 for $Z_{PLF}$ = 18, 19. This
ratio exceeds by 11$\%$  the $N/Z$ of the initial neutron-rich
projectile ($^{48}$Ca $N/Z=1.4$). 
We observe the following features:
\begin{itemize}
\item an increase of the mean value ($\langle A_{PLF} \rangle$) 
and width ($\sigma$) of the distributions as the considered $Z_{PLF}$ 
increases. For small $Z_{PLF}$ no significant differences in the mean and width 
of the isotopic distributions for the four systems are observed, while both 
increase significantly with increasing $Z_{PLF}$ for the n-rich projectile
(\nuc{48}{Ca}). The effect is most pronounced for $Z_{PLF}=Z_{proj}$.
\item No dependence of the isotopic distributions on the target is observed for the 
\nuc{40}{Ca} projectile, whereas a shift toward higher $\langle A_{PLF} \rangle$
and $\sigma$ is observed when moving from the \nuc{40}{Ca} to \nuc{48}{Ca}
target. 
\end{itemize}
The shift of the mean values and the broadening of the distributions
as the n-richness of the system increases can be understood in terms of
available neutrons. The evolution of the behaviour observed
as $Z_{PLF}$ increases reflects the evolution of the products with
the centrality of the collision, therefore with the interaction time
and the flux of nucleons exchange. 

The analysis of these results following the lines of refs. \cite{Ono04,Col13}
is in progress and should provide new information on the symmetry energy term of 
EOS of nuclear matter. 

%%%%%%%%%%%%%%%%%%%%%%%%%%%%%%%%%%%%%%%%%%%%%%%%%%%%%%%%%%%%%%%%%%%%%%%%%%%%%%%%%%%%%
\subsection{Using a new approach for solving the Boltzmann-Langevin equation 
\label{sect63}}
%%%%%%%%%%%%%%%%%%%%%%%%%%%%%%%%%%%%%%%%%%%%%%%%%%%%%%%%%%%%%%%%%%%%%%%%%%%%%%%%%%%%%
\begin{figure}[b!]\begin{center}
\includegraphics[angle=0, width=\columnwidth]{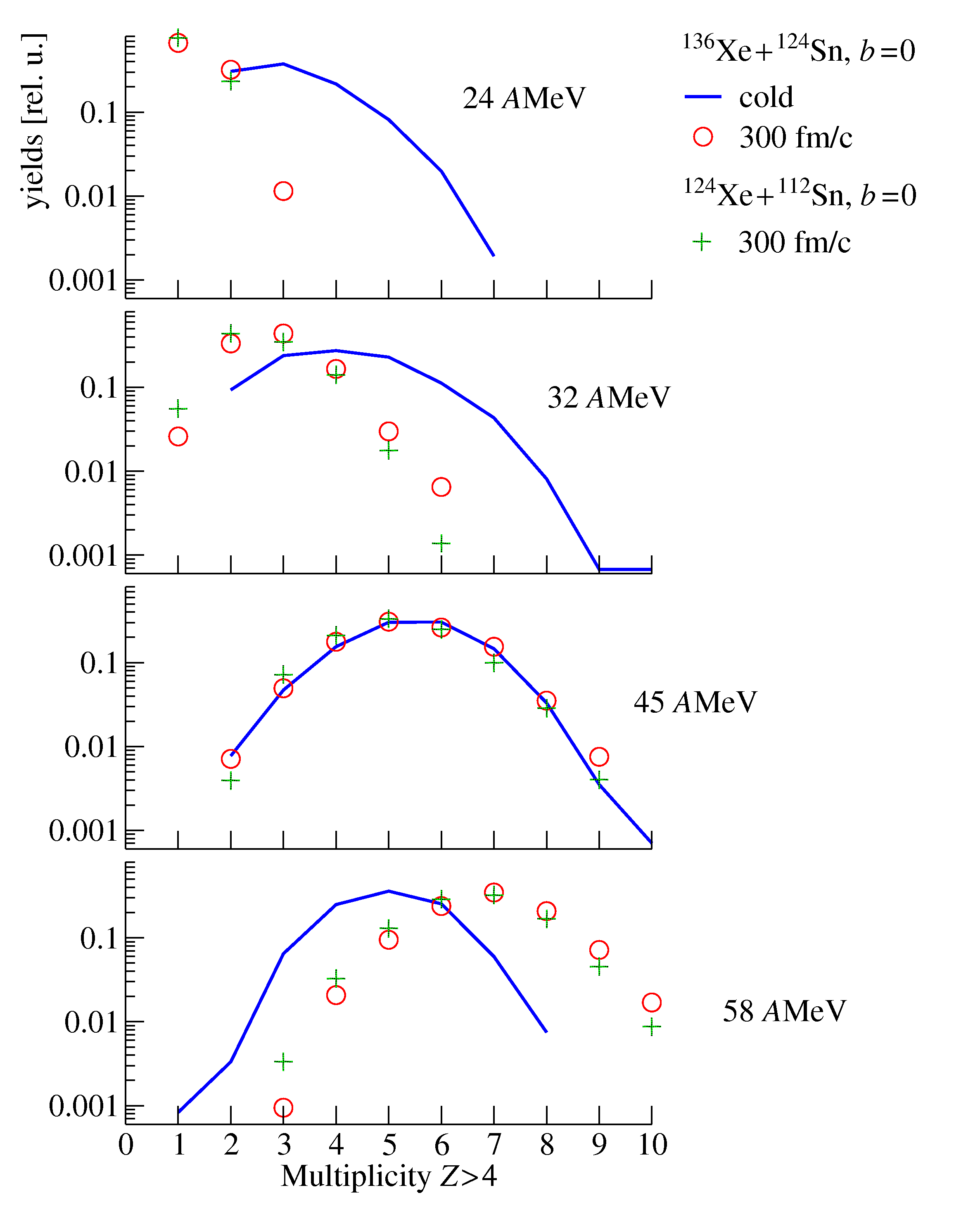}
\end{center}\caption
{Multiplicity distribution of fragments with $Z\!>\!4$ obtained 
for four incident energies with BLOB,  for $^{136}$Xe$+^{124}$Sn 
and $^{124}$Xe$+^{112}$Sn reactions, at $300$ fm/c for $b\!=\!0$.
The first system is also studied after secondary decay.}
\label{fig_BLOB}
\end{figure}
%%%%%%%%%%%%%%%%%%%%%%%%%%%%%%%%%%%%%%%%%%%%%%%%%%%%%%%%%%%%%%%%%%%%%%%%%%%%
As we discussed in section \ref{sect3}, for central collisions the theoretical 
approach SMF which applies rather efficiently at larger bombarding energies 
(Xe$+$Sn at 45~\AM{}), does not extend to lower energies.
From a microscopic point of view, the drop in fragment multiplicity in 
proximity of the low-energy multifragmentation threshold may result from 
a complex interplay between spinodal instabilities, which tend to fragment 
the system in several portions of comparable size, and the coalescence effect
of the mean-field, which tends to revert the system to a compact shape.
This picture requires first of all that the many-body description 
(fragment observables) at low bombarding energies should be done without
sacrifying the one-body properties, like the spinodal behaviour and, 
secondly, that the fermionic model is sufficiently stable
(against turning into a Boltzmann statistics), which is a fundamental 
requirement for following relatively long-time processes like coalescence effects.
A suited approach to extend to lower energies was therefore developed with the
aim of keeping the same accuracy in describing isospin and transport properties
(i.e. the isovector-dependent mean field) along a larger incident-energy 
interval which includes also the low-energy threshold of multifragmentation.
\emph{Such extension required the solution of the Boltzmann-Langevin (BL) 
equation in three dimensions in order to describe the fluctuation phenomenology 
in full phase space.}

A discussion on the problems which may be encountered in introducing a Langevin 
term in a transport model is presented by Chapelle et al. in ref.~\cite{Cha92},
and a solution was the method introduced by Rizzo et al. in ref.~\cite{Rizz08}, 
tested in nuclear matter. 
The recommendation from the above references was to pay special attention to 
the Pauli blocking;  this becomes therefore the direction followed for building 
the Boltzmann-Langevin One Body model, BLOB~\cite{Nap13,NapIWM11}.
With the BLOB simulations, the multifragmentation process can be followed down 
to low energy  where the system 
explores instabilities against density fluctuations when low densities are 
attained, and a characteristic spinodal behaviour stands out.
The fact that, in this situation, the amplitude of the unstable modes grows 
according to the specific dispersion relation associated with the employed 
mean-field interaction ensures that the Langevin term is implemented 
satisfactorily.

With the BLOB model 
head-on collisions for the system \sys{136}{Xe}{124}{Sn} can be described 
quantitatively down to low energy, as shown in fig.~\ref{fig_BLOB} for 
four incident energies. At 24~\AM{}, large fragment multiplicities are present.
At 32~\AM{}, conversely to SMF simulations, one observes a spinodal behaviour.
45~\AM{} corresponds with the multifragmentation regime 
and to the situation when secondary decay does not modify the primary 
distribution of fragments, and 58~\AM{} is at the fading side of the
spinodal multifragmentation process.

To compare more quantitatively with the data discussed in section~\ref{sect3},
we selected collisions with impact parameters smaller than 4~fm from 
simulated events (asy-stiff EOS) for the system \sys{136}{Xe}{124}{Sn} 
at 32~\AM{}. After de-excitation of the hot fragments one finds 
$\langle M_{frag} \rangle$=4.5, $\langle Z_{lcp} \rangle$=36.0, 
$\langle Z_{b5} \rangle$=60.6. These preliminary low statistics results 
can be compared to the experimental values \\
$\langle M_{frag} \rangle$= 5.5, 
$\langle Z_{lcp} \rangle$ =26.7, $\langle Z_{b5} \rangle$= 71.0.
While underestimation of the fragment multiplicity and of $Z_{b5}$ are of
the same order of magnitude as that observed at 45~\AM{} with SMF,
the value of $Z_{lcp}$  is in better agreement with the data, because
preequilibrium emission is reduced.

Thus the BLOB model extends the description of multifragmentation to lower 
energies than the SMF model. It is due to the following mechanism:
fluctuations have the effect of reducing the fraction of energy spent in the 
emission of light-particles (the so-called preequilibrium). More energy is 
therefore available for feeding the development of inhomogeneities 
and, at the same time, for imparting a large kinetic energy to them: 
the dynamics results more explosive and is able to drive the fragment
separation and to produce events of large fragment multiplicity.
Such modelling approach results then in a more correct description of 
the velocity profile of fragments. It should allow to get information on 
the symmetry energy stiffness by looking, for instance, at the evolution 
of the average N/Z of light fragments \emph{vs} their centre of mass 
kinetic energy, as proposed in~\cite{Colo08,T47Gag10}.

%%%%%%%%%%%%%%%%%%%%%%%%%%%%%%%%%%%%%%%%%%%%%%%%%%%%%%%%%%%%%%%%%%%%%%%%%%%%%%%%%%%%%
\section{Foreseen studies \label{sect7}}
%%%%%%%%%%%%%%%%%%%%%%%%%%%%%%%%%%%%%%%%%%%%%%%%%%%%%%%%%%%%%%%%%%%%%%%%%%%%%%%%%%%%%

The advent of radioactive intense beam facilities altogether with the 
development of new-generation $4\pi$ arrays~\cite{FAZIA13} will allow to 
deeply investigate 
the physics of isospin in the  forthcoming years. Indeed, a large panoply
of n-rich and p-rich beams will become available with bombarding energies 
from the Coulomb barrier up to few hundreds of MeV per nucleon. 
From a perspective point of view~\cite{LoI06,Eur03}, the exploration of the 
isospin degree of freedom through heavy-ion induced reactions will bring up the 
possibility of studying :
\begin{itemize}
\item limiting temperatures in hot N/Z asymmetric nuclear systems: Coulomb versus 
isovector instabilities,
\item isospin dependence of nuclear level densities in warm nuclei: 
continuation of INDRA-VAMOS program (section 5.1.2),
\item isospin diffusion and migration through dissipative collisions,
\item isospin dependence of the nuclear phase diagram: phase transitions and 
coexistence lines, spinodal (mechanical) versus isovector (chemical) 
instabilities, hot versus ``cold'' multifragmentation,
\item the nuclear symmetry energy directly from fragment isotopic distributions: 
isoscaling as detailed below. 
\end{itemize}

In the mean time, coupling the FAZIA demonstrator with INDRA, we foresee the
following investigation. 
We performed theoretical study in the 
framework of a $3D$ Lattice-Gas Model, (\emph{LGM})~\cite{Leh10}. In this 
simulation, we have implemented an \emph{isocalar+isovector+Coulomb} Hamiltonian 
in order to evaluate the combined effects of nuclear, isospin and Coulomb 
interactions concerning the fragment production of excited nuclear systems produced 
in heavy-ion reactions in the Fermi energy domain. Details concerning the 
implementation and results from this model can be found in \cite{Leh10,T45Leh09}.   

%%%%%%%%%%%%%%%%%%%%%%%%%%%%%%%%%%%%%%%%%%%%%%%%%%%%%%%%%%%%%%%%%%%%%%%%%%%%%%%%%%%%%
\subsection{Density functional \label{sect71}}
%%%%%%%%%%%%%%%%%%%%%%%%%%%%%%%%%%%%%%%%%%%%%%%%%%%%%%%%%%%%%%%%%%%%%%%%%%%%%%%%%%%%%

To relate the system properties to a macroscopic description in
term of \emph{density functional}, we have implemented a \emph{Liquid Drop}
parametrization for the internal energy with coefficients which are 
density-dependent. Thus, the internal energy $E_{int}$ of the nuclear system 
at density $\rho$ and isospin content $\delta=(N-Z)/A$ can be expressed as:
 
\begin{eqnarray}
\lefteqn{E^{LD}_{int}(\rho,\delta)=} \nonumber \\
& &[ a_v(\rho)+c_{sym}^v(\rho) {\delta}^2 ] A +
[ a_s(\rho) + c_{sym}^s(\rho) {\delta}^2 ] A^{2/3} \nonumber \\
& & + \alpha_c(\rho) Z^2
\label{df}
\end{eqnarray}
where $a_v$, $c^v_{sym}$, $c^s_{sym}$, $a_s$ and $\alpha_c$ are respectively 
the volume, symmetry (volume and surface terms), surface and Coulomb 
coefficients. The label $LD$ stands here for \emph{Liquid Drop}. 

%%%%%%%%%%%%%%%%%%%%%%%%%%%%%%%%%%%%%%%%%%%%%%%%%%%%%%%%%%%%%%%%%%%%%%%%%%%%%%
\begin{figure}[htb]
\centering \includegraphics[width=0.8\columnwidth]{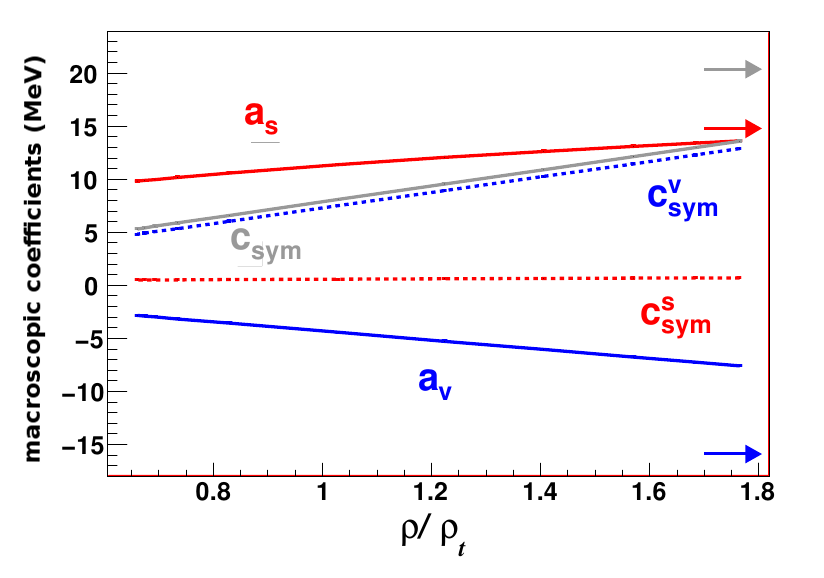}
\caption{Density dependence of the $LD$ coefficients. The arrows indicate the 
values obtained for $c_{sym}^{v}$  $a_s$ and $a_v$ at saturation density/zero 
temperature. From~\cite{Leh09}.} \label{LG_fonctionnelle}
\end{figure}
%%%%%%%%%%%%%%%%%%%%%%%%%%%%%%%%%%%%%%%%%%%%%%%%%%%%%%%%%%%%%%%%%%%%%%%%%%%%%%%%

Fig. \ref{LG_fonctionnelle} shows the density dependence of
the $LD$ coefficients.  We notice that all coefficients
are decreasing with the density and that the surface term of the symmetry energy
$c_{sym}^s$ is here equal to zero; the main contribution for the symmetry term is
indeed the volume coefficient $c_{sym}^v$ in this implementation
of \emph{LGM}~\cite{Leh10}, as also found in AMD simulations (see
section~\ref{sect62}).

%%%%%%%%%%%%%%%%%%%%%%%%%%%%%%%%%%%%%%%%%%%%%%%%%%%%%%%%%%%%%%%%%%%%%%%%%%%%%%%%
\subsection{Isospin dependence of the symmetry energy \label{sect72}}
\begin{figure}[htb]
\includegraphics[width=\columnwidth]{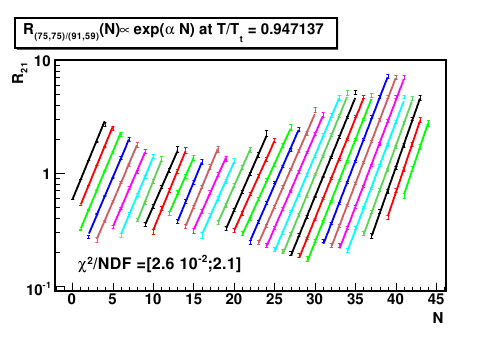}
\caption{(Colour on line) Yield ratio $R_21(Z,N)$ as a function of the number 
of neutrons $N$ for nuclei with $Z=1-36$ for \emph{LGM} simulations.  
From \cite{Leh09}.}
\label{LG_isoscaling}
\end{figure}

%%%%%%%%%%%%%%%%%%%%%%%%%%%%%%%%%%%%%%%%%%%%%%%%%%%%%%%%%%%%%%%%%%%%%%%%%%%%%%%%
 Isotopic scaling has been experimentally observed several years ago by the 
\emph{MSU} group~\cite{Xu00,Tsa01}, and also found later in various
experimental data~\cite{Soul03,R4-Ger04,I56-Lef05}. It is built upon the ratio of
production yields obtained from two sets of nuclear reactions having the same 
total mass $A$ but different isospin ratios $N/Z$ and $N'/Z'$ (which are hereafter
labeled $(1)$ and $(2)$). More precisely, this is an isotopic scaling law, 
abbreviated as \emph{isocaling}, which expresses the ratio $R_{21}(Z,N)$ between 
the production yields $Y_i(Z,N)$, for both reactions, of a fragment defined 
by $Z$ protons and $N$ neutrons as : 
\begin{equation}
R_{21}(Z,N)=\frac{Y_2(Z,N)}{Y_1(Z,N)} \propto \exp( \beta Z + \alpha N )
\label{isoscaling}
\end{equation}
where $\alpha$ and $\beta$ are the two isoscaling parameters related respectively 
to the number of neutrons $N$ and the number of protons $Z$. This scaling is
characterized by \emph{parallel straight lines} in a logarithmic plot of $R_{Z,N}$ 
as a function of $N$ or $Z$ whatever is the fragment. The slope of the straight 
lines is either $\alpha$ or $\beta$ depending on the representation, for
instance it is $\alpha$ on Fig. \ref{LG_isoscaling} obtained from 
\emph{LGM} simulations. 

%%%%%%%%%%%%%%%%%%%%%%%%%%%%%%%%%%%%%%%%%%%%%%%%%%%%%%%%%%%%%%%%%%%%%%%%%%%%%%
%%%%%%%%%%%%%%%%%%%%%%%%%%%%%%%%%%%%%%%%%%%%%%%%%%%%%%%%%%%%%%%%%%%%%%%%%%%%%%%%

 It has been proposed to relate this experimental evidence for the 
 \emph{isocaling} to fundamental properties of the nuclear equation of state, 
 namely the symmetry energy and its coefficient $c_{sym}$ (see Eq.~\ref{df}). 
 Different parameterizations have been derived
either from macroscopic or microscopic approaches. 
   
%%%%%%%%%%%%%%%%%%%%%%%%%%%%%%%%%%%%%%%%%%%%%%%%%%%%%%%%%%%%%%%%%%%%%%%%%%%%%%%%
\subsubsection{Macroscopic approach \label{sect721}}   
%%%%%%%%%%%%%%%%%%%%%%%%%%%%%%%%%%%%%%%%%%%%%%%%%%%%%%%%%%%%%%%%%%%%%%%%%%%%%%%%
 The first parametrization is based on the macroscopic approach from 
statistical models, in the grand-canonical framework~\cite{Botv02}. In this 
specific case, the \emph{isocaling} parameter $\alpha$ is related to the symmetry 
energy coefficient $c_{sym}$ through :  
 \begin{equation}
 c_{sym}(Z) = \frac{ \alpha(Z) T}{4(Z_2^2/A_2^2-Z_1^2/A_1^2)}
 \label{csym_evap}
 \end{equation}
where $Z_1$,$A_1$ and $Z_2$,$A_2$ are the total atomic numbers and masses of 
the two distinct isospin systems, $T$ is the temperature, and $\alpha(Z)$ is the 
\emph{isocaling} coefficient for fragment with charge $Z$. We can see the 
results of Eq.~\ref{csym_evap} plotted as dashed curves on Fig.~\ref{LG_csym}. 
Each curve corresponds to a given $Z$, varying from $Z=2$ to $Z=7$. 
We restrict here the \emph{isoscaling} to light fragments as it is usually done 
in experimental conditions~\cite{Xu00,Tsang01,Soul03,R4-Ger04,I56-Lef05}. 
The values are far away from the true ones displayed by the symbols as it has
been already observed in~\cite{Das08}. Several reasons could explain this large 
discrepancy. Eq.~\ref{csym_evap} is not exact and has been derived from a 
macroscopic framework where the many-body correlations are supposed to be 
exhausted entirely by clusterization~\cite{Leh09}. This is a rather crude 
approximation because it appears to be strongly affected by conservation law, 
combinatorial effects~\cite{Cha08} but also by secondary decay
effects~\cite{Tsa06}. At last, Eq.~\ref{csym_evap} is actually related to the
symmetry \emph{free energy} which should correspond to the symmetry energy 
only when $T\rightarrow0$. 

%%%%%%%%%%%%%%%%%%%%%%%%%%%%%%%%%%%%%%%%%%%%%%%%%%%%%%%%%%%%%%%%%%%%%%%%%%%%%%%%
\subsubsection{Microscopic approach \label{sect722}}
%%%%%%%%%%%%%%%%%%%%%%%%%%%%%%%%%%%%%%%%%%%%%%%%%%%%%%%%%%%%%%%%%%%%%%%%%%%%%%%%

%%%%%%%%%%%%%%%%%%%%%%%%%%%%%%%%%%%%%%%%%%%%%%%%%%%%%%%%%%%%%%%%%%%%%%%%%%%%%%
\begin{figure}[htb]
\includegraphics[width=\columnwidth]{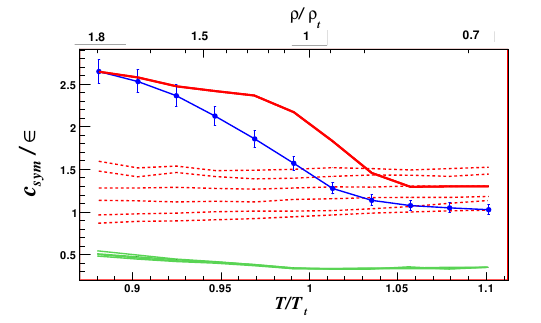}
\caption{(Colour on line) Various evaluations of the symmetry energy coefficients 
(see text) compared to the true one (symbols). Green dashed curves:
Eq.~\ref{csym_evap}; red dotted curves: Eq.~\ref{csym_frag}; thick red line
is for the largest cluster. From \cite{Leh09}.}
\label{LG_csym}
\end{figure}
%%%%%%%%%%%%%%%%%%%%%%%%%%%%%%%%%%%%%%%%%%%%%%%%%%%%%%%%%%%%%%%%%%%%%%%%%%%%%%%%
A second parametrization, derived from the quantal microscopic model 
\emph{AMD} in the fragmentation regime~\cite{Ono04}, has been also established:  

\begin{equation}
 c_{sym}(Z) = \frac{ \alpha(Z) T}{4(Z^2/<A_1>^2-Z^2/<A_2>^2)}
 \label{csym_frag}
 \end{equation}

 where $<A_i>$ is here the average mass number for a fragment with charge $Z$ in 
 the reaction $(i)$. The results are displayed by the dotted light grey (flat) 
 curves in fig.\ref{LG_csym} for the same range of atomic number ($Z=2-7$) as
 previously. We notice that the 
 values are quite correct for the low density/high temperature domain, below
(above) the density/temperature transition. This is indeed expected since 
Eq.~\ref{csym_frag} has been derived for the fragmentation regime only. 
For the high density/low temperature domain, the agreement is not anymore observed.
  This reflects again the difficulty to extract reliable values for $c_{sym}$ 
from the  isoscaling parameters extracted by looking at light fragments ($Z<8$) 
where combinatorial  and secondary decay effects may blur the signal. 
  
%%%%%%%%%%%%%%%%%%%%%%%%%%%%%%%%%%%%%%%%%%%%%%%%%%%%%%%%%%%%%%%%%%%%%%%%%%%%%%%%
\subsection{\emph{Isoscaling} from the largest fragment \label{sect73}}
%%%%%%%%%%%%%%%%%%%%%%%%%%%%%%%%%%%%%%%%%%%%%%%%%%%%%%%%%%%%%%%%%%%%%%%%%%%%%%%%
 When analysing the results from \emph{LGM}, we notice that the mass distribution
 includes a large percolating cluster which contains
most information on the thermodynamics~\cite{Ric01,Leh10}. Therefore,
the isotopic distribution of the \emph{largest cluster} may be more sensitive 
to the symmetry energy of the fragmenting system. We then apply Eq.~\ref{csym_frag}
to the largest cluster in the event, and this is plotted in Fig.~\ref{LG_csym}
 as a continuous thick red line. We observe a better,
though still qualitative, agreement with the true value from the model. 
This is particularly interesting since the largest fragment is by definition
bound  in \emph{LGM}~\cite{Ric01}; this means that this is the \emph{final} 
distribution of the largest fragment which is here analyzed, avoiding thus 
the secondary decay problem pointed out previously. In the vivid perspective 
of obtaining experimental measurements of isotopic distributions for heavy 
fragments from experiments coupling recoil spectrometers
with large acceptance arrays like MARS with NIMROD at Texas $A\&M$~\cite{Soul06} or
VAMOS with INDRA at GANIL~\cite{ChbIWM07}, or new generation $4\pi$ arrays like
FAZIA~\cite{FAZIA13} associated with radioactive beam facilities, we could 
expect to get unprecedented results concerning the estimation of the symmetry 
energy and its associated
density/temperature dependence in heavy-ion collisions. 

%%%%%%%%%%%%%%%%%%%%%%%%%%%%%%%%%%%%%%%%%%%%%%%%%%%%%%%%%%%%%%%%%%%%%%%%%%
\section{General conclusion \label{sect8}}
%%%%%%%%%%%%%%%%%%%%%%%%%%%%%%%%%%%%%%%%%%%%%%%%%%%%%%%%%%%%%%%%%%%%%%%%%%
In this review paper, we gathered all the isospin effects observed in
experiments performed with INDRA, essentially with stable beams, which
limits the explored N/Z range (1 - 1.5). Nuclear reactions were studied 
from the barrier to  100~\AM{}. The Fermi energy domain is well suited for 
constraining the symmetry energy term of the EOS as the mean field still 
plays an important role. Subsaturation and moderate
suprasaturation densities are explored during these collisions.

For central collisions, we did not observe isospin influence on the stopping
between 30 and 100~\AM{} in the studied isospin range. On the other
hand isospin effects are visible in various properties of the
multifragmentation process, but did not allow to constrain the symmetry 
energy when compared to the SMF transport code. The BLOB model now allows to 
extend the data-model comparisons to lower energies, which will be done in
the near future.

Investigating isospin diffusion as a function of dissipation on a rather
large range of impact parameters, we showed that, within the SMF framework,
the potential part of the symmetry energy linearly increases with density.
We also demonstrated that isospin equilibration occurred for semi-central
collisions with 8-9 \AM{} available energy. These two properties were
observed for both a symmetric and a very asymmetric system.

In a foreseeable future we will couple the FAZIA demonstrator with INDRA.
FAZIA should perform isotopic identification for Z up to 25. Tke knowledge of
the mass of rather heavy fragments on a large solid angle is expected to 
permit a step forward in constraining the density dependence of the symmetry 
energy and to give better insight into isospin effects in nuclear reactions. 

%%%%%%%%%%%%%%%%%%%%%%%%%%%%%%%%%%%%%%%%%%%%%%%%%%%%%%%%%%%%%%%%%%%%%%%%%%%%
\appendix
\section{Experimental selections with INDRA \label{sect2}}
%%%%%%%%%%%%%%%%%%%%%%%%%%%%%%%%%%%%%%%%%%%%%%%%%%%%%%%%
Nucleus-nucleus collisions are classified through their violence, 
which reflects the impact parameter. INDRA is particularly efficient for 
 central to semiperipheral collisions at intermediate energy. Around
and above the Fermi energy most of the collisions end-up with two big nuclei, 
remnants of projectiles (QP) and target (QT). A fraction of collisions 
also presents a copious and fast emission of particles and light 
fragments with velocities intermediate between those of the QP and QT. 
For central collisions topology selectors allow to isolate quasifusion 
reactions whereas impact parameter selectors are able to isolate the 
most central collisions. Depending on collisions
and required studies, different selections are used (see for 
example~\cite{I28-Fra01,Bor08}).
As INDRA does not detect neutrons, selections do not lead to N/Z
reconstructions of nuclear sources.

%%%%%%%%%%%%%%%%%%%%%%%%%%%%%%%%%%%%%%%%%%%%%%%%%%%%%%%%%%%%%%%%%%%%%%%%%%%%
\subsection{Event selection for central collisions\label{sect21}}
%%%%%%%%%%%%%%%%%%%%%%%%%%%%%%%%%%%%%%%%%%%%%%%%%%%%%%%%%%%%%%%%%%%%%%%%%%%%
A two step procedure is used to select central collisions. We shall
distinguish, for the second step, quasifusion (QF) sources selections, for
which one uses a topology selector (events with isotropic shape in velocity 
space), from global selections of the most central collisions which need an
impact parameter selector. At intermediate energies there are large
fluctuations in the exit channel associated to a given impact parameter, as
shown in stochastic transport models~\cite{Colon98}, and such a distinction
makes sense.

The first step consists in keeping the events for which a quasi-complete 
detection of the reaction products has been achieved. Significant fractions:
$\geq 77-80 \%$ of the charge of the system, $Z_{sys} = Z_{proj} + Z_{targ}$,
is required to be measured for every event.
In the second step we need  a topology selector in order to select
well defined nuclear source issued from quasifusion reactions.
The flow angle ($\Theta_{flow}$) 
selection~\cite{Cug83,Lec94,I28-Fra01} is largely used.
This global variable is defined as the
angle between the beam axis and the preferred direction of emitted matter
in each event. It is determined by the energy tensor calculated  from 
fragment ($Z \geq 5$) momenta in the reaction centre of mass. Quasifusion events 
have no memory of the entrance channel and should be isotropic while binary 
dissipative collisions are focused at small $\Theta_{flow}$. Thus, by 
selecting only large flow angles, fusion events can be well isolated.
The minimum flow angle chosen is 60$^\circ$ for collisions in
the incident energy range 30-50~\AM{}.
%In this paper we chose 60$^\circ$ for all energies, to get enough
%statistics (at least 30000 events) without degrading the properties of a
%compact fused source. 
The present selection corresponds to measured cross sections decreasing 
from 30-40 to 20-25~mb when the incident energy goes from 30 to 50~\AM{}
for Xe + Sn collisions. 
By taking into account detection efficiency and biases due to the selection 
(quasi-complete events and flow angle selection) the total cross section for 
the formation of compact fused systems is estimated to decrease from 200-250 to
80-150~mb between 30 and 50~\AM{}~\cite{I39-Hud03,I77-Gag12}. 

In the experiment leading to the results presented in sections \ref{sect3}
and \ref{sect412} we collected a very large number of events; it thus became
possible to perform more severe selections, in order to extrapolate different 
variables to the values corresponding to a perfect detection of charges, 
Z$_{tot}$=104. We proceed as follows: we considered, for each system, three 
batches of events with the conditions Z$_{tot} \geq$80, 90 and 95. We 
calculated average multiplicities of LCP, $M_{lcp}$ and of fragments with 
lower limits Z=3 ($M_{f3}$) and Z=5 ($M_{frag}$). We also determined the total 
charge bound in LCP or fragments, $Z_{lcp}$, $Z_{b3}$, $Z_{b5}$ respectively. 
It appeared that all these variables linearly evolve when plotted \emph{vs} 
the average values of Z$_{tot}$. These linear relations were used to 
extrapolate the multiplicities and Z$_{bound}$ values to what they would be for 
a perfect detection. The consistency of the procedure was verified by  
considering the sum $Z_{lcp} + Z_{b3}$; it was in all cases equal to 
104$\pm$0.01, except for \sys{136}{Xe}{112}{Sn} at 32~\AM{} where it 
amounts to 104.3.

\begin{figure}[htb]
\includegraphics[width=\columnwidth]{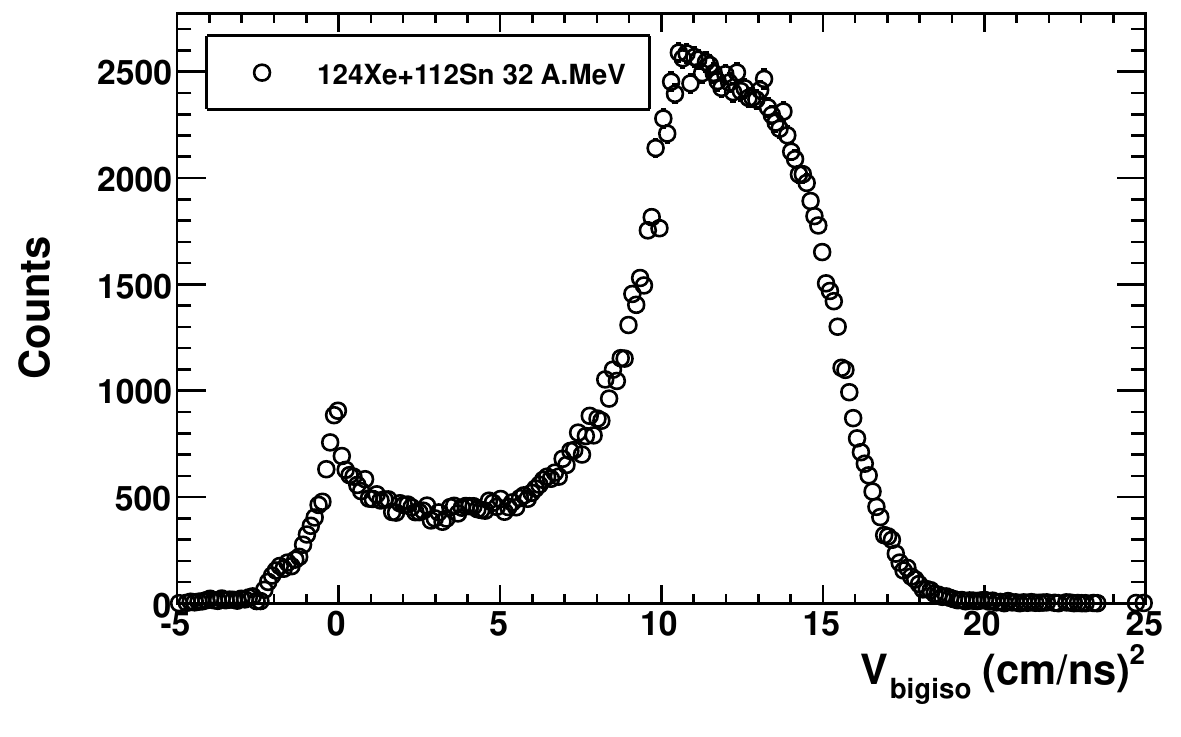}
\caption{ $V_{bigiso}$ distributions for a Xe + Sn reaction at 
32~\AM{} incident energy. From~\cite{T49Kab13}.}
\label{Vbigiso}
\end{figure}

More recently
another topology selector was proposed to select QF sources. For symmetric
systems, taking advantage of the excellent quality detection of INDRA
in the centre of mass forward hemisphere, the quantity
$V_{bigiso}$ = $V^2_{par}$ - 0.5$V^2_{per}$ for the heaviest fragment
was calculated~\cite{T49Kab13}; $V_{par}$ and $V_{per}$ refer to centre of
mass velocity components parallel and perpendicular to the beam direction. 
A value of
$V_{bigiso}$ close to zero is expected for QF events, for which the velocity
distribution of the heaviest fragment must be isotropic. An example of
such a selection is shown in figure~\ref{Vbigiso} for different Xe + Sn
reactions at 32~\AM{} incident energy. 
%%%%%%%%%%%%%%%%%%%%%%%%%%%%%%%%%%%%%%%%%%%%%%%%%%
%%%%%%%%%%%%%%%%%%%%%%%%%%%%%%%%%%%%%%%%%%%%%%%%%%%%%%%

When one wants to select the most central collisions as for stopping studies
an impact parameter selector must be preferred. With such a selection mean
properties of a well defined class of collisions can be extracted.
Since quantities involving both longitudinal and tranverse directions will
be used to evaluate the stopping, it is not suitable to use vector variables
such as, for instance the transverse energy~\cite{I10-Luk97,I17-Pla99} which could
introduce autocorrelations. A scalar variable is desirable an a natural choice
is the total multiplicity of detected charged products, $N_{ch}$.

%%%%%%%%%%%%%%%%%%%%%%%%%%%%%%%%%%%%%%%%%%%%%%%%%%%%%%%%%%%%%%%%%%%%%%%%%%%
\subsection{Selection of quasiprojectiles and impact parameter evaluators\label{sect22}}
%%%%%%%%%%%%%%%%%%%%%%%%%%%%%%%%%%%%%%%%%%%%%%%%%%%%%%%%%%%%%%%%%%%%%%%%%%%
As previously, a two step selection is applied.
A first and simple selection requires that the total detected charge 
amounts to at least 90 (80)\% of the charge of the projectile (in the
forward part of the centre of mass).

A further selection must be done to select the ``quasi-projectile''.
 We do not intend to isolate a ``source'', but rather to select a
forward region in phase space where the detected products have a small
probability to result from emission by the quasi-target. For symmetric systems, 
it is done
by a cut at the centre-of-mass velocity; we only keep the forward part
 for which, as previously said, detection is excellent.
For asymmetric systems like Ni+Au,  
the target being more than three times heavier than the projectile, some
particles from the target would be kept.
Thus the cut was made at the nucleon-nucleon velocity.
The quasi-projectile selection only
keeps particles and fragments with a parallel velocity higher than
the nucleon-nucleon velocity. 

So, in the following \textit{we call ``quasi-projectile'' (QP) the ensemble of 
charged products which have a velocity higher than the centre of mass 
(nucleon-nucleon for asymmrtric systems) velocity, without
prejudice on the equilibration of any degree of freedom of the ensemble so
defined.}

%%%%%%%%%%%%%%%%%%%%%%%%%%%%%%%%%%%%%%%%%%%%%%%%%%
\begin{figure}[htb]
\includegraphics[width=\columnwidth]{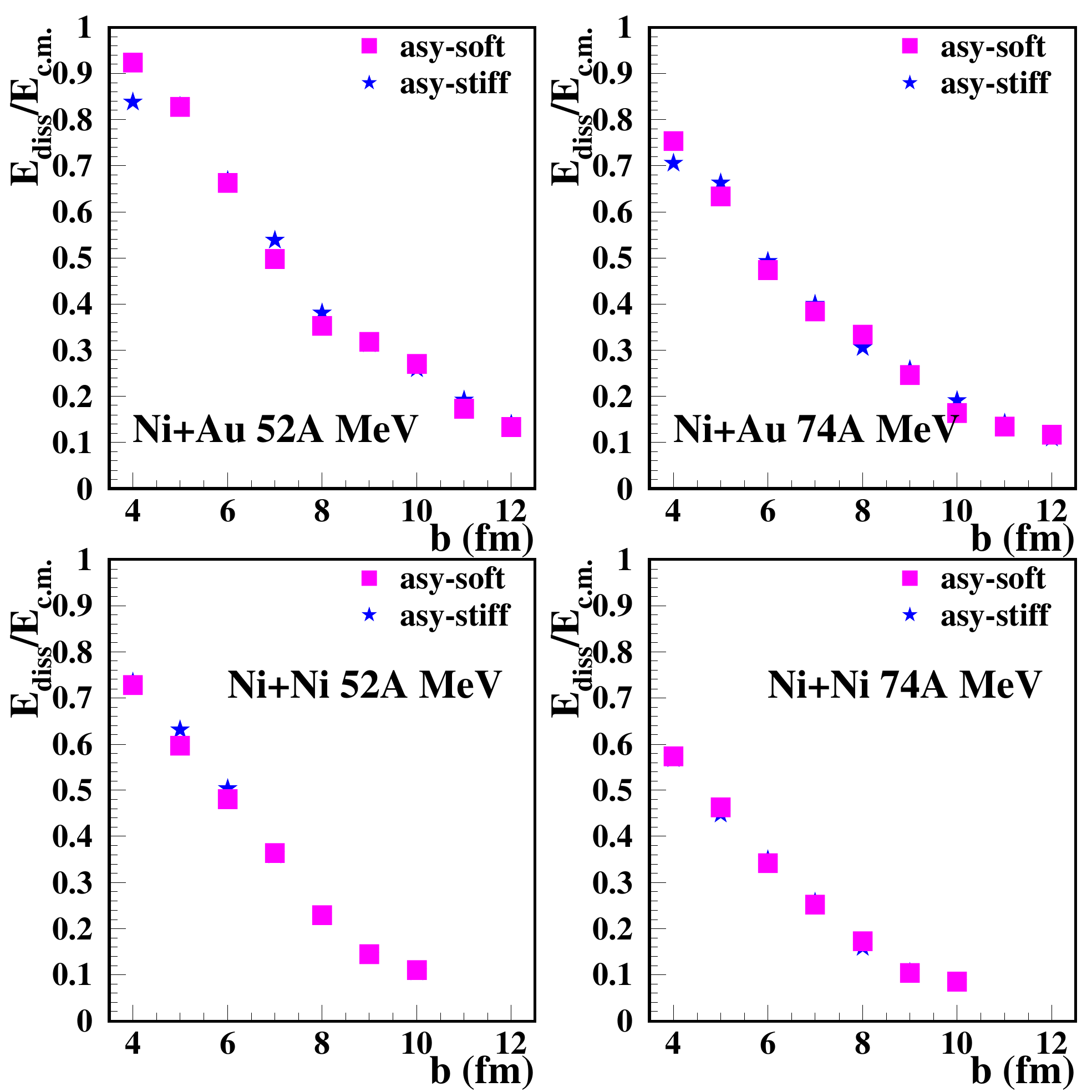}
\caption{(Colour on line) Correlation between $E_{diss}$/$E_{cm}$ and
the true impact parameter for two reactions and two incident energies. 
Stars (squares) correspond to SMF simulations (see section~\ref{sect31}) 
with an asy-stiff (asy-soft) symmetry energy. From~\cite{Gal09}.
}
\label{fig:16}
\end{figure}
%%%%%%%%%%%%%%%%%%%%%%%%%%%%%%%%%%%%%%%%%%%%%%%%%%%%%%%
To study the evolution of observables and to compare
with theoretical simulations, impact parameter evaluators have to be used.
First of all, as early mentioned, we have chosen the  transverse
energy of light charged particles (LCP, Z=1,2), $E_{tr12}$, 
but now corresponding to the forward
part of the centre of mass. This observable is quite well correlated with
the impact parameter, especially for peripheral and semiperipheral
collisions, and well suited to the data because the INDRA array is very
efficient (85\%) for LCP's.

We have also sorted the events as a function
of the dissipated energy, calculated in
a binary hypothesis, with the following assumptions: \\
i) the quasi-projectile velocity, $V_{QP}^{rec}$, is equal to
the measured velocity of the single fragment, or reconstructed from the velocity of
all the fragments it contains,

ii) the relative velocity between the quasi-projectile and the quasi-target
is determined as if the collision was purely binary, without mass exchange:
\begin{equation} \label{eq:vrel}
V_{rel}=V_{QP}^{rec} \times \frac{A_{tot}}{A_{target}}
\end{equation} 
and thus the total dissipated energy reads:
\begin{equation} \label{eq:Eexc}
E_{diss}=E_{c.m.}-\frac{1}{2}\mu V_{rel}^2 ,
\end{equation}
with $\mu$ the initial reduced mass.
It is demonstrated in~\cite{Yan03,Pia06} that the velocity of the QP is a 
good parameter for following the dissipated energy, except in very 
peripheral collisions, due to trigger conditions. Moreover, it is shown
in figure~\ref{fig:16} that in transport model  simulations (SMF, see
section~\ref{sect31}) $E_{diss}$ gives a good measure of 
the impact parameter when using the sorting of events as
previously defined.

%\bibliography{\repi mftp,\repi articles_indra,\repi theses_indra,\repi conf,%
%\repi revues,\repi RetD,\repi articles_reverse}
%% Faire un premier passage latex. ensuite taper ``bibtex fich'', qui
% utilise le fichier fich.aux cree. puis refaire un latex - ou deux

\ifx\mcitethebibliography\mciteundefinedmacro
\PackageError{epja_cplus.bst}{mciteplus.sty has not been loaded}
{This bibstyle requires the use of the mciteplus package.}\fi

\end{document}